\documentclass[11pt]{article}

\usepackage[scaled]{helvet}

\usepackage{amsmath,bm}
\usepackage{graphicx,pdflscape}
\usepackage{color}
\usepackage[toc,page]{appendix}
\usepackage{simpler-wick}

\usepackage{color,subfigure}

\usepackage{hyperref}
\hypersetup{
    colorlinks,
    citecolor=red,
    filecolor=cyan,
    linkcolor=blue,
   urlcolor=magenta
 }

\usepackage{xparse}
\ExplSyntaxOn
\NewDocumentCommand{\mref}{m}{\quinn_mref:n {#1}}
\seq_new:N \l_quinn_mref_seq
\cs_new:Npn \quinn_mref:n #1
 {
  \seq_set_split:Nnn \l_quinn_mref_seq { , } { #1 }
  \seq_pop_right:NN \l_quinn_mref_seq \l_tmpa_tl
  ( 
  \seq_map_inline:Nn \l_quinn_mref_seq
    { \ref{##1},\nobreakspace } 
  \exp_args:NV \ref \l_tmpa_tl 
  ) 
 }
\ExplSyntaxOff

\newcommand{\disp}[1]{Eq.~\mref{#1}}
\newcommand{\dispmany}[1]{Eqs.~\mref{#1}}

\newcommand{\figdisp}[1]{Fig.~\mref{#1}}
\newcommand{\refdisp}[1]{Ref.~(\cite{#1})}

\newcommand{\lessim} {\ {\raise-.5ex\hbox{$\buildrel<\over\sim$}}\ }

\newcommand{\gssim}{\ {\raise-.5ex\hbox{$\buildrel>\over\sim$}}\ }

\newcommand{\si}{\sigma}
\newcommand{\sib}{\bar{\sigma}}
\newcommand{\tJ}{\ $t$-$J$ \ }
\newcommand{\nn}{\nonumber}

\renewcommand{\emph}{\textit}

\newcommand{\iden}{ {\bf 1}}

\newcommand{\sr}{ \textcolor{black}}

\usepackage{hyperref}

\newcommand{\beq}{\begin{eqnarray}}
\newcommand{\eeq}{\end{eqnarray}}
\newcommand{\barray}{\begin{eqnarray}}
\newcommand{\earray}{\end{eqnarray}}

\newcommand{\bra}{\langle\langle}
\newcommand{\ket}{\rangle\rangle}
\newcommand{\half}{\frac{1}{2}}
\newcommand{\chem}{{\bm \mu}}
\newcommand{\A}{{\cal A}}
\newcommand{\G}{{\cal G}}
\newcommand{\F}{{\cal F}}
\newcommand{\f}{{\bm f}}

\newcommand{\GH}{{\bf g}}
\newcommand{\GHI}{\GH^{-1}}

\newcommand{\wt}{\widetilde}
\newcommand{\wh}{\widehat}

\newcommand{\wmu}{\widetilde{\mu}}
\newcommand{\mbstar}{\frac{m}{m^*}}
\newcommand{\ppsi}{\Psi(\mu_0)}
\newcommand{\xip}{ \varepsilon_p - \mu_0}
\newcommand{\sign}{\mbox{sign}}

\newcommand{\J}{{\cal J}}
\newcommand{\Hunc}{H_{\mbox{unc-tJ}}}

\graphicspath{{Figures/}}

\begin{document}

\title{  Extremely Correlated  Superconductors }

\author{ B Sriram Shastry$^{1}$\footnote{sriram@physics.ucsc.edu}  \\
\small \em $^{1}$Physics Department, University of California, Santa Cruz, CA, 95064 \\}
\date{September 2, 2021}

\maketitle

\abstract{   
Superconductivity  in the \tJ model is studied by  extending  the recently introduced extremely correlated fermi liquid theory. Exact  equations for the Greens functions are obtained by  generalizing Gor'kov's equations to include extremely strong local repulsion between electrons of opposite spin. These equation are expanded  in a parameter $\lambda$ representing the fraction of double occupancy, and the lowest order equations are further simplified near $T_c$,  resulting in an approximate integral equation for the superconducting gap. The condition for $T_c$ is studied using a model spectral function embodying a reduced quasiparticle weight $Z$  near half-filling, yielding an approximate analytical formula for $T_c$. 
 This formula is evaluated   using  parameters representative of  single layer High-$T_c$ systems. In a narrow  range of electron densities that is necessarily  separated from the Mott-Hubbard insulator at half filling, we find   a typical  $T_c$$\sim$$10^2$K. 
}

\vspace{0.25in}


\vspace{0.25in}

\section{Introduction \label{Intro}}

The single band \tJ model \disp{tJ},\cite{tJname,Zhang-Rice}, and the closely related strong coupling Hubbard model have attracted much attention  in recent years. In large part the interest is  due to the potential relevance of these models in describing the phenomenon of  High $T_c$ superconductivity, discovered in cuprate materials in 1987 \cite{Bednorz} and later, in other materials.  These models lead to  a single sheet of the fermi surface, and are specified by fixing  the band hopping $t$ and the  exchange energy $J$ for the \tJ model,  or equivalently  $4 t^2/U$ for the strong coupling ($U\gg t$) Hubbard model, \sr{where the interaction is given by  $V_{Hub.}= U \sum_i n_{i \downarrow} n_{i \uparrow}$}. The exotic possibility of
superconductivity arising from such inherently repulsive systems, is surprising from a theoretical perspective, and also challenging. \sr{ Significant theoretical work using a variety of tools
on the strong coupling Hubbard model and the extremely strong coupling  \tJ models  }\cite{Hirsch,Anderson, BZA,Kotliar,Gros,Yokoyama,Mele,Mohit,Rice,Giamarchi,Ogata,Num-5,TVR}  has given useful insights into the role of strong correlations in cuprate superconductivity.  However  given the non-triviality of the theoretical task of \sr{ analytically  solving these models,  progress in that direction has been slow.} 

 \sr{ In this work we  extend  the  extremely correlated fermi liquid theory (ECFL)  \cite{ECFL,ECFL-AOP-1} recently formulated to overcome the  analytical   difficulties of the strong coupling models, to include   superconducting type broken symmetry.  Upon cooling the normal metallic state,   a superconducting instability is expected to  arise, and our main goal in this study  is to determine  the conditions for the occurrence of this state, and to provide its detailed  description.}
 
\sr{ In order to motivate these calculations of the superconducting state, it is useful to summarize the main features of ECFL theory  as applied to the normal (non-superconducting) state so far. We provide   a broad overview next, further details can be found in \refdisp{ECFL,ECFL-AOP-1}. }

\sr{  The methodology developed in this theory starts with  exact functional differential equations for the various Greens function, obtained using the Tomonaga-Schwinger approach of external potentials. These equations incorporate the modification of the  anti-commutation relations between the fermion operators  due to Gutzwiller projection (see \disp{anticommutator}). 
 While providing a formally exact starting point for us, these equations are not yet amenable to systematic approximations. 
 The core difficulty is that an   additional set of  terms arise from this modified  non-canonical anticommutator structure \disp{anticommutator}.  These non-canonical terms multiply the  most singular term in the equation, namely the Dirac delta function (originating in the time derivative of the time ordering $\Theta$ functions in the Greens functions). For  an explicit example, note     the $\gamma$ term multiplying the delta function in \disp{G0XY}.}

\sr{ In order to make progress, we therefore need to go beyond the 
established framework of Tomonaga-Schwinger. The first development in ECFL is that the above inconvenient feature of a non-canonical coefficient of the delta function,  is eliminated by factoring the Greens function into two parts, the auxiliary Greens function $\GH$ and the caparison function $\wmu$ (see \disp{factorize} and the discussion in the text following it).  The auxiliary Greens function $\GH$ now  satisfies a canonical equation (as in \disp{new-1} by ignoring the term involving $\f$), while  the  caparison function $\wmu$ accounts for the non-canonical nature of the original equation  (as in \disp{new-2}).  This factorization process and the resulting equations are  exact. }

\sr{  As the next  development, we introduce a parameter $\lambda$ in the range $0\leq \lambda \leq 1$ into these exact equations. Setting $\lambda$$=$$0$ gives the uncorrelated system, while $\lambda$$=$$1$ gives the exact equations of the strongly correlated system.
 The $\lambda$ parameter has a formal  similarity to the expansion parameter $\frac{1}{2 S}$  used in the Dyson-Maleev (or Holstein-Primakoff) formulations \cite{DM,HP} of the spin-wave theory of magnets. The magnetic  models  involve spin operators  satisfying the   SU(2) (angular momentum) Lie algebra. 
  They  can be approached  using  different  strategies. On the one hand we may think of spins as canonical bosons with a constraint on their occupation number $n^b_{i}$ at any site $i$, namely    $n^b_i=0,1,\ldots 2S$. This constraint can be implemented using  a repulsive interaction between bosons $U n^b_i (n^b_i-1)\ldots (n^b_i-2S)$, and finally letting $U$$\to$$ \infty$. This bosonic Hubbard model  is difficult to solve, since the large energy scale  $U$ makes the use of perturbation theory impractical. On the other hand we can employ the Dyson-Maleev (or Holstein-Primakoff) non-linear mappings to bosons, and expand the relevant Heisenberg equations of motion in a series  in $\frac{1}{2 S}$. This gives  an efficient way of solving the models to considerable precision at fairly low  orders in $\frac{1}{2 S}$. This latter method  is parallel  to the $\lambda$ expansion employed here,  since the modified anticommutators  \disp{anticommutator} also yield a (non-canonical) Lie algebra.This analogy is discussed further in \refdisp{ECFL-AOP-1} (Sec. 6). In  a different setting, the parameter $\lambda$ can also be related to  the fraction of doubly occupied states \cite{ECFL-Formalism} (see Appendix. A). }

 \sr{ The parameter $\lambda$ serves two important and related objectives. Firstly it provides a continuous path between the uncorrelated and  the fully correlated system equations.
 Since $0\leq \lambda\leq 1$, dialing it up  from $0$  does not involve invoking a large energy scale,  unlike for example,  dialing  up $U$ in the Hubbard model. This (isothermal) continuity enables the ECFL method to   retain the ideal (i.e. non-interacting)  fermi surface volume  at low T. This ideal volume is expected for weakly interacting fermi systems from the   Luttinger-Ward  perturbative arguments \cite{Luttinger},  and  importantly,  survives the transition to  extremely strongly  correlated regions, as argued  recently using non-perturbative arguments \cite{Shastry-FS-Volume}. Lastly,    the ideal volume is also seen in photoemission studies of  overdoped and optimally doped cuprate superconductors in the normal state \cite{ARPES-Luttinger-Ward}, which provide a useful  starting point for  our study.}

\sr{ The second aspect of $\lambda$ is that it can be used to organize a systematic power series  expansion, analogous in spirit to the skeleton graph expansion of Dyson \cite{Dyson} in perturbative theories.   This    $\lambda$ expansion    can be carried out order by order, leading to a set of successive  equations that are amenable to numerical study. A question might arise, whether a low order calculation in this expansion can capture the  strongly correlated limit.
For answering this,  it is useful to examine the results for the  $d=\infty$ Hubbard model at $U=\infty$, where numerically
exact results are available from the dynamical mean field theory \cite{DMFT}.  The $\lambda$ expansion to  ${\cal O}(\lambda^2)$ is compared  with the exact numerical result from the dynamical mean field theory \cite{Rok},
in Fig.~(6) of \refdisp{Shastry-Perepelitsky}.     This shows that the calculated quasiparticle weight $Z$  vanishes  upon approaching  a density of $1$-particle per site, i.e. half filling. This vanishing  is a  hallmark of  the  strong correlation limit, where the Mott-Hubbard insulating state is realized. In the above $d=\infty$ study,  and also in the case of the 2-dimensional \tJ model \cite{Resistivity,NJP,PRB-MaiShastry}, the $\lambda$ expansion describes an extremely correlated Fermi liquid state, characterized by  a small quasiparticle weight that vanishes near the Mott-Hubbard insulator, accompanied by  a rich set of low energy scales located above the (strongly suppressed)  effective fermi temperature. The  ${\cal O}({\lambda}^2)$ equations for the normal state have  been applied  to calculations of  the asymmetric photoemission lines\cite{Gweon,NJP,PRB-MaiShastry}, and most recently the
calculation of the almost T-linear resistivity in single layer cuprates\cite{Resistivity}. }

\sr{ In this paper we extend the above formalism to the case where superconducting order emerges at low temperatures. This requires a non-trivial generalization to the superconducting state of the various steps of the ECFL theory highlighted  above. In a similar fashion to the normal state, we first  obtain exact equations  for the normal and anomalous    Greens functions for the \tJ model. These equations   generalize Gor'kov's equations for  BCS type weak coupling superconductivity\cite{Gorkov} by including the effect of extremely strong local repulsion between electrons.
 These equations  are studied further using a  specific  decomposition of the Greens functions into two pieces  (see \disp{factorize}).   This step is
  followed by a systematic expansion in  a parameter $\lambda$.  This leads to an  set of equations \disp{new-1,new-2,new-3}, iterating these in $\lambda$ to all orders constitutes the  exact answer. In the  present work, we  perform  a leading order calculation. }

 \sr{ In order to obtain explicit  results, \disp{new-1,new-2,new-3} are further simplified near $T_c$ where the order parameter is small, leading to simplified versions of these in    \disp{new-4,new-5,new-6}.  These are treated to ${\cal O}({\lambda}^2)$,   and the lowest order  condition for $T_c$ is formulated in \disp{Tc1}. In summary  \disp{Tc1} is the leading order term near $T_c$,  within the  $\lambda$ expansion, and  constitutes  an important  formal result of the present work. In principle it should be possible to find further systematic equations to higher order, and also to extend the results for $T\ll T_c$ following the procedure laid out here. In this work we are content to  study this first set in  detail.   The transition temperature is given from \disp{Tc1},  which is expressed in terms of the electronic Greens function, renormalized by strong correlations. In this renormalization  the short ranged Hubbard-Gutzwiller terms are dominant, and the pairing energy causing the instability,  is provided by the much smaller  exchange energy $J$. This equation  exhibits both a   tendency towards an insulating state due to a diminished quasiparticle weight, and a tendency towards superconductivity due to the exchange term $J$. Their competing tendencies  play out in \disp{Tc1} and the closely related \disp{Tc2}. These equations determine whether superconductivity is found at all,  and further identifies  the model parameters that promote it. When the  superconducting state is found, they also provides an estimate of  the range of densities and temperatures which favor it.  }

The conditions \disp{Tc1,Tc2}  are evaluated using a simple  phenomenological electronic  spectral function,  modeling   strong correlations near half filling in terms of a density dependent quasiparticle weight $Z$ and a wide  background. This model has the advantage of leading to an explicit analytical formula for $T_c$, in terms of the various parameters of the \tJ model, thus allowing for a thorough understanding of the role of different parameters on the result. Evaluating this expression we find that the model supports a d-wave superconducting phase consistent with data \cite{d-wave-experiment,d-wave-review},  located  away from half filling. The $T_c$ is found to be typically $\sim$$10^2$K, i.e. an order of magnitude smaller than that of the  model \disp{unctJ} \sr{where the sole difference from the \tJ model is that short ranged Hubbard-Gutzwiller type correlations are ignored}, in a range of densities determined by the band parameters. The temperature-density phase diagram has the form of a tapered tower \figdisp{Fig-1-Tc}. A smooth  dome structure reported   in cuprates, is replaced here by  a somewhat narrow  density range and an exaggerated height near the peak.   The location of the peak can be varied by choosing the hopping parameters, but always remains well-separated from the insulating limit.

The paper is organized as follows. In Section (\ref{Preliminaries}) we define the \tJ Hamiltonian,  express it in terms of the correlated fermionic operators, and outline the method of external potentials employed to generate the exact dynamical equations for   the electron Greens function $\G$ and the Gor'kov anomalous Greens function $\F$. In Section (\ref{lambdaexpansion})  the equation is expanded in $\lambda$ and further simplified near $T_c$.    In Section (\ref{EquationforTc}) the condition for $T_c$ is evaluated using a model spectral function.  This section contains expressions that   involve only the electronic   spectral function, and  might be  directly accessible to readers who are more interested in the concrete results.
In Section (\ref{Conclusions}) we conclude with  a discussion of the results.

\section{Theoretical Preliminaries \label{Preliminaries}}

The \tJ Hamiltonian \cite{tJname,Zhang-Rice} is 
\beq
H_{\mbox{ tJ}}&=&H_{\mbox{t}}+H_{\mbox{J}} \label{tJ}  \\
H_{\mbox t}&=& -\sum_{ij \si} t_{ij} \wt{c}^\dagger_{i \si} \wt{c}_{j \si} - \chem \sum_i n_i  \nn \\
H_{\mbox{J}}&=& \half \sum_{ij} J_{ij} (\vec{S}_i.\vec{S}_j - \frac{n_i n_j}{4}) \nn \eeq
\sr{where $t_{ij}$ are the band hopping matrix elements detailed below, $J_{ij}$ the nearest neighbor exchange and $\chem$ the chemical potential,  }
with the density operator  $n_i = \sum_\si \wt{c}^\dagger_{i \si}\wt{c}_{i \si}$, and spin density operator  $S_i^\alpha =\half \sum_{\si \si'} \wt{c}^\dagger_{i \si} \tau^\alpha_{\si \si'}\wt{c}_{i \si'}$, $\tau^\alpha$ is a Pauli matrix and 
 the correlated fermi destruction operator $\wt{c}_i$  is found from the plain (i.e. canonical or unprojected) operators $c_i$, by sandwiching it between two Gutzwiller projection operators $\wt{c}_{i \si}=P_G c_{i \si} P_G$, \sr{where $P_G\equiv \prod_j(1-n_{j \uparrow} n_{j \downarrow})$ \cite{Gutzwiller}. It acts by eliminating all states with double occupancy in the state space.} The creation operators follow by taking their hermitean conjugate. The physical meaning of this sandwiching process is that the fermi operators act within the subspace where projector $P_G$  enforces single occupancy at each site. \sr{The \tJ model may be  obtained by taking the large $U$ limit of the Hubbard model \cite{tJname}. It has also been argued \cite{Zhang-Rice} to be  the low energy effective Hamiltonian for an underlying  three-band model, describing the copper oxygen lattice of the cuprate superconductors, where it is found  by eliminating high energy states of the model.  }

  In the following work we will also find it useful to study the model 
 \beq
 \Hunc= -\sum_{ij} t_{ij} {c}^\dagger_{i \si} {c}_{j \si} - \chem \sum_i n_i + \half \sum_{ij} J_{ij} (\vec{S}_i.\vec{S}_j - \frac{n_i n_j}{4}). \label{unctJ} 
 \eeq  
 \sr{We may view it as an {\em uncorrelated \tJ model} in contrast to the correlated version  \disp{tJ}, here the ultra strong short ranged Hubbard-Gutzwiller   correlations with $U \gg  \mbox{max} \{ |t_{ij}|\}$ are turned off, while   the  relatively weak     exchange term $J\ll \mbox{max} \{ |t_{ij}|\}$ is retained.  All  operators that appear in \disp{unctJ}, including the density and spin, are defined by the same expression as \disp{tJ}  but with the  unprojected fermion operators $c_{i \si}, c^\dagger_{i \si}$'s. In this model  the exchange term, which  is usually viewed as the mechanism for antiferromagnetism,   doubles up to   play the  role of a superconducting  pairing potential. This fruitful  observation of  Anderson, Baskaran and Zou\cite{Anderson,BZA} follows from viewing the interaction in the crossed or Cooper channel.  It    is paralleled  in  our  discussion later (see paragraph below \disp{exchange-2}), where the exchange term, after a rearrangement  amounting to a crossed channel,  leads to a mean Cooper pair expectation in \disp{exchange-3}. Its superconducting solution, found by standard BCS-Gor'kov meanfield theory, is presented below (see \dispmany{Tcuncorr,g}), and  serves as a useful reference point in the study of the strongly  correlated \tJ model. }

 It is  convenient for our calculations  to use the operators invented by Hubbard \refdisp{Hubbard,Ovchinnikov}  to represent this projection process.  \sr{  \refdisp{Edward}
  (Sec.8) discusses  the origin of difficulties  of the early work employing the Hubbard operators, in reproducing  the  Luttinger-Ward Fermi surface volume at low temperatures. In contrast   the  present ECFL  formalism achieves this goal successfully, using continuity with the Fermi gas and the $\lambda$ expansion described in \cite{ECFL, Edward} and below.}  We denote
\beq
\wt{c}^\dagger_{i \si} \leftrightarrow X_i^{ \si 0}, \; \wt{c}_{i \si} \leftrightarrow X_i^{0 \si}, \; \wt{c}^\dagger_{i \si} \wt{c}_{i \si'} \leftrightarrow X_i^{\si \si'}.
\eeq
These operators satisfy the following fundamental anti-commutation relations and their adjoints:
\beq
\{X_i^{0 \si_i}, X_j^{0 \si_j } \}&=&0\\
\{X_i^{0 \si_i}, X_j^{\si_j 0} \}&=& \delta_{ij} \left( \delta_{\si_i \si_j}- \si_i \si_j X_i^{\sib_i \sib_j} \right), \;\;\sib=-\si. \label{anticommutator}
\eeq
In physical terms, for a given site index $i$ and  with $\{a,b\} \in$ $\{0, \uparrow, \downarrow\}$ limited to the  three allowed initial and final states of the projected Hilbert space, the symbol $X_i^{ab}$ represents an operator  representing  all allowed matrix elements. To yield the correct fermion antisymmetry, the creation operator $X_i^{\si_i, 0}$     anti-commutes  with creation or destruction operators  at {\em different} sites with any spin.  In terms of these operators we can rewrite 
\beq
H_t&=& - \sum_{ij \, \si} t_{ij} X_i^{\si 0} X_{j}^{0 \si} -\chem \sum_{i \, \si} X_i^{\si \si}\label{Ht}\\
H_J&=&= - \frac{1}{4} \sum_{ij \, \si_i \si_j} J_{ij} \si_i \si_j X_i^{\si_i \si_j} X_j^{\sib_i \sib_j}. \label{HJ}
\eeq
\sr{In the following we employ a convenient repeated internal  spin summation convention. We shall follow the convention that in an equation defining  any object, often (but not always) indexed by  external  spin indices, all the {\em internal
 and repeated} spin indices are to be summed over. As an example, we could drop the explicit summation over spins in  \disp{Ht,HJ}, but not in \disp{anticommutator} where $\si_i,\si_j$ are external spin indices that appear on the left hand side.  We also use a repeated internal  site index below. }

In order to calculate the Greens functions for this model, we add an imaginary time $\tau$ dependent external potential (or source term) ${\cal A}$ to the definition of thermal averages. The expectation of an arbitrary observable  $Q(\tau_1,\ldots)$,  composed  e.g. of a product of several (imaginary) time ordered  Heisenberg picture operators, is written  in  the notation 
\beq
\bra Q(\tau_1,\ldots) \ket=  Tr\; P_\beta \; T_\tau \{ e^{-{\cal A}} Q(\tau_1,\ldots) \}. \label{Bracket}
\eeq
Here $T_\tau$ is the time-ordering operator,  an external potential term ${\cal A}= \int_0^\beta d\tau \A(\tau)$, and   $P_\beta = e^{-\beta H} /{Tr \left( e^{-\beta H} T_\tau e^{-{\cal A}}\right)} $ is the Boltzmann weight factor including ${\cal A}$. Here $\A(\tau)  $ is a sum of two terms, ${\cal A}_{\rho}(\tau)$ involving a  density-spin dependent external potential ${\cal V}$, and $  {\cal A}_C(\tau)$ involving ${\cal J}$ (${\cal J}^*$)  Cooper pair generating (destroying) external potentials. These  are given by
\beq
{\cal A}_{\rho}(\tau)& =& \sum_i   {\cal V}_i^{\si_i \si_j}(\tau) X_i^{\si_i \si_j}(\tau) \nn \\ 
{\cal A}_{C}(\tau)& =&\half \sum_{ij}  \left( {\cal J}^*_{ j\si_j i\si_i}(\tau) X_i^{0\si_i}(\tau) X_j^{0 \si_j}(\tau) +{\cal J}_{i\si_i j\si_j}(\tau) X_i^{\si_i 0}(\tau) X_j^{ \si_j 0}(\tau) \right), \nn \\
 \label{Sources}
\eeq
\sr{where  the repeated internal spin convention implies summing over $\si_i,\si_j$, and}
where we require the antisymmetry ${\cal J}_{i \si_i; j \si_j}=-{\cal J}_{j \si_j; i \si_i}$ and likewise for ${\cal J}^*$.
The external potentials ${\cal J}, {\cal J}^*$ in \disp{Sources}  couple to operators that add and remove  Cooper pairs of correlated electrons, and are essential to describe the superconducting phase. At the end of the calculations, the external potentials are switched off, so that the average in \disp{Bracket} reduces to the standard thermal average.
Tomonaga\cite{Tomonaga}  in 1946 and Schwinger\cite{Schwinger}  in 1948   (TS) pioneered the use of such external potentials \cite{Dyson,Martin}. We next illustrate
 this technique for the present problem.
 
\subsection{Using  external potentials \label{ExternalPotentials}}
The advantage of introducing these external potential  ( or ``sources'') is that   we can take the (functional) derivatives of Greens function with respect to the added external  potentials in order  to generate higher order Greens functions. If we abbreviate  the external term   as ${\cal A}= \sum_i {\cal U}_j(\tau) V_j(\tau)$, where ${\cal U}_j(\tau)$ is one of the above 
 c-number  potential, and
$V_j(\tau)$ is the corresponding operator in the imaginary-time Heisenberg picture, and $Q_i(\tau)$ an arbitrary observable, straightforward differentiation leads to  the TS identity
\beq
Tr P_\beta  T_\tau \{ e^{-{\cal A}} Q_i(\tau') V_j(\tau) \}=\bra  Q_i(\tau')\ket \; \bra V_j(\tau) \ket - \frac{\delta}{\delta {\cal U}_i(\tau)} \bra  Q_i(\tau')\ket  \label{TS}
\eeq
This important  identity can be found by taking the functional derivative of
\disp{Bracket} with respect to ${\cal U}_j(\tau)$ (see e.g. \refdisp{ECFL-Formalism} Eq.~(18)), and    is now illustrated  with various choices of the external potential.

From \disp{TS} we note the frequently used result
\beq
\bra \si_i \si_j X_i^{\sib_i \sib_j}(\tau) Q(\tau') \ket = \left(\gamma_{\si_i\si_j}(i \tau) - {\cal D}_{\si_i\si_j}(i \tau) \right) \bra Q(\tau')\ket \label{CalD}
\eeq
where
\beq
\gamma_{\si_i \si_j}(i, \tau)&=& \si_i \si_j \bra X_i^{\sib_i \sib_j}(\tau) \ket \nn \\
 {\cal D}_{\si_i \si_j}(i, \tau))&=&\si_i \si_j \frac{\delta}{\delta {\cal V}_i^{\sib_i \sib_j}(\tau)}, \label{Def-gamma}
\eeq


The singlet Cooper pair operator is 
\beq
\left(X_i^{ 0 \uparrow} X_j^{0 \downarrow }- X_i^{0 \downarrow } X_j^{ 0 \uparrow}\right)= \si X_i^{0 \si} X_j^{0 \sib},
\eeq
where summation over  $\si$  is implied on the right hand side, and its Hermitean conjugate
\beq
-\left(X_i^{ \uparrow  0} X_j^{ \downarrow  0}- X_i^{ \downarrow  0 } X_j^{  \uparrow  0}\right)= \sib X_i^{ \si 0} X_j^{ \sib 0}.
\eeq
We define the (singlet) Cooper pair  correlation functions at time $\tau$ as
\beq
C_{ij}(\tau)= \bra \si X_i^{0\si}(\tau) X_j^{0 \sib}(\tau)\ket  \label{Cij}\\
C^*_{ij}(\tau)= \bra \sib X_i^{\si 0}(\tau) X_j^{ \sib 0}(\tau)\ket \label{Cijstar},
\eeq
\sr{where $\si$ is summed over.}
We note that $C^*_{ij}$  equals the complex conjugate of $C_{ij}$ only after  the external potentials are finally  turned off, but not so in the intermediate steps. 

The basic equation \disp{TS} for the Cooper pair operators for an arbitrary operator $Q$ are
\beq
\frac{\delta}{\delta \J^*_{i \si_i j \si_j}(\tau)}\bra Q\ket= \bra  X_j^{0\si_j}(\tau) X_i^{0 \si_i}(\tau)\ket \bra Q\ket -\bra  X_j^{0\si_j}(\tau) X_i^{0 \si_i}(\tau) Q\ket \\
\frac{\delta}{\delta \J_{i \si_i j \si_j}(\tau)}\bra Q\ket= \bra   X_i^{0 \si_i}(\tau) X_j^{0\si_j}(\tau)\ket \bra Q\ket -\bra   X_i^{0 \si_i}(\tau) X_j^{0\si_j}(\tau) Q\ket 
\eeq
From these relations the Cooper-pair correlations can be found by summing over the spins 
\beq
\bra \si X_i^{0\si}(\tau) X_j^{0 \sib}(\tau) Q  \ket = \left[ C_{ij}(\tau)-{\cal K}_{ij}(\tau) \right] \bra Q \ket \label{Kappa} \\ 
\bra \sib X_i^{\si 0}(\tau) X_j^{ \sib 0}(\tau) Q  \ket = \left[ C^*_{ij}(\tau)-{\cal K}^*_{ij}(\tau) \right] \bra Q \ket \label{Kappastar} 
\eeq
where
\beq
{\mathcal K}_{ij}(\tau)=\sib  \frac{\delta}{\delta{\cal J}^*_{i \si; j \sib}(\tau)} \\
{\cal K}^*_{ij} =  \sib  \frac{\delta}{\delta{\cal J}_{i \si; j \sib}(\tau)},
\eeq
\sr{where $\si$ is summed over.}

 
  \subsection{Greens functions and their dynamical  equations \label{EOMs}}
 We are interested in the electron Greens function \sr{ (see e.g. \refdisp{ECFL-Formalism} Eq.~(17)) expressed compactly by}
 \beq
\G_{i \si_i j \si_f}(\tau,\tau')= - \bra X_i^{0 \si_i}(\tau) X_j^{\si_f 0 }(\tau') \ket, \label{eq-G}
\eeq
where the \sr{Dyson  time ordering $T_\tau$ and the external potential factor $e^{-{\cal A}}$ are}   included in the definition of the brackets \disp{Bracket}.
To describe the superconductor, following  Gor'kov \cite{Gorkov} we define  the anomalous Greens function :
\beq
\F_{i \si_i j \si_f}(\tau,\tau')=  \sib_i  \bra X_i^{ \sib_i 0}(\tau) X_j^{\si_f 0 }(\tau') \ket \label{F} 
\eeq
where $\sib\equiv -\si$,   \sr{and as in \disp{eq-G}, the Dyson  time ordering $T_\tau$ and the external potential factor $e^{-{\cal A}}$ are}   included in the definition of the brackets \disp{Bracket}

We note that the Cooper pair correlation functions \disp{Cijstar}, which plays a crucial  role in defining the order parameter of the superconductor, can be expressed in terms of the anomalous Greens function using
\beq
C^*_{ij}(\tau)&=& \F_{i \si j \si}(\tau,\tau), \label{cstarf}
\eeq
where $\si$ is to be summed over, as per the convention used.
We will  also need the equal time correlation of creation operators ${C}_{ij}(\tau)$ \disp{Cij}.
It is straightforward to show that when the external potentials ${\cal A}$ are switched off, this object is independent of $\tau$ and can be obtained by complex conjugation of $C_{ij}^*$. It is possible to add another anomalous Greens function with two destruction operators as in \disp{F},
corresponding to Nambu's generalization of Gor'kov's work. In the present context it  adds little to the calculation and is avoided by taking the complex conjugate of ${C}^*_{ij}$ to evaluate ${C}_{ij}$.

\subsubsection{Greens function $\G$ \label{EOM-G}}
The equations for the Greens functions follow quite easily from the Heisenberg equations, followed by the use of the identity \disp{TS}, and has been discussed extensively by us earlier. There is one new feature, concerning  an alternate treatment of the  $H_J$ (exchange) term, necessary for describing superconductivity described below. \sr{In this section we make use of the internal repeated site index summation convention quite extensively.}

Taking the $\tau$ derivative of $\G$ we obtain
\beq
\partial_{\tau} \bra X_i^{0 \si_i}(\tau) X_f^{\si_f 0}(\tau') \ket =\delta(\tau-\tau') \delta_{if} (\delta_{\si_i \si_f} - \gamma_{\si_i \si_f}(i \tau) ) \nn \\
 + \bra[H_t+H_J+\A(\tau),X_i^{0 \si_i}(\tau)]\;\; X_f^{\si_f 0}(\tau')\ket
\eeq

We work on the terms on the right hand side.  At time $\tau$ we note
\beq
[H_t+\A_{\rho},X_i^{0 \si_i}]= \chem  X_i^{0 \si_i} - {\cal V}_i^{\si_i \si_j} X_i^{0 \si_j}
+t_{ij} (\delta_{\si_i \si_j}- \si_i \si_j X_i^{\sib_i \sib_j}) X_j^{0 \si_j},
\eeq
\sr{where the repeated internal indices $\si_j$ and $j$ are summed over}.
From this basic commutator,  using \disp{TS}, \disp{CalD} and the definitions \disp{Def-gamma} we obtain
 \beq
   \bra[H_t+\A_{\rho} (\tau),X_i^{0 \si_i}(\tau)]\;\; X_f^{\si_f 0}(\tau')\ket =\left( \chem \delta_{\si_i \si} - {\cal V}_i^{\si_i \si} \right)  \bra X_i^{0 \si}(\tau) X_f^{\si_f 0}(\tau')\ket \nn \\
   + t_{ij} \bra X_j^{0 \si_i}(\tau) X_f^{\si_f 0}(\tau')\ket - t_{ij} ( \gamma_{\si_i \si}(i, \tau)- {\cal D}_{\si_i \si}(i, \tau))\bra X_j^{0 \si}(\tau)  X_f^{\si_f 0}(\tau')\ket,
  \eeq
  \sr{where the repeated spin index $\si$, and the site index $j$  are  summed over, while $\si_i,\si_f$ and site indices $i,f$ are held fixed.}


For the exchange term
 \beq
 [H_J,X_i^{0  \si_i }]&=&  \half J_{ij}   \si_i \si \;   X_i^{ 0 \si }   X_j^{ \sib_i \sib }  \label{exchange-1} \\
  &=& - \half J_{ij} \si_i X_j^{ \sib_i 0} \left(X_i^{ 0 \uparrow} X_j^{0 \downarrow }- X_i^{0 \downarrow } X_j^{ 0 \uparrow}\right), \label{exchange-2}
  \eeq
  \sr{where the repeated internal indices $\si$ and $j$ are summed over}.
In order to obtain \disp{exchange-2} from   \disp{exchange-1},  we used  $X_j^{\sib_i \sib} =X_j^{ \sib_i 0} X_j^{0 \sib }$ and anticommuted the equal time operators $X_i^{ 0 \si } X_j^{\sib_i 0} $  into $-  X_j^{\sib_i 0} X_i^{ 0 \si } $, followed by \sr{an explicit sum} over $\si$.
This subtle step is  essential  for obtaining the superconducting phase, as discussed (para following \disp{unctJ}) in the Introduction, since the role of exchange in promoting Cooper pairs manifests itself here.  
Using \disp{Kappa} we find
 \beq
   \bra [H_J,X_i^{0 \si_i}(\tau)] X_f^{\si_f 0}(\tau')\ket =-\half J_{ij} \si_i \left(C_{ij}(\tau^+)-{\cal K}_{ij}(\tau^+)  \right) \bra X_j^{\sib_i 0}(\tau)  X_f^{\si_f 0}(\tau')\ket, \nn \\ \label{exchange-3}
\eeq 
\sr{where the repeated internal index $j$ is summed over, and with $\eta$ is a positive infinitesimal we indicate here and elsewhere $ \tau^+ \equiv \tau+ \eta $ and $ \tau^- \equiv \tau- \eta $  }.

In treating this term  we could have proceeded differently by sticking to \disp{exchange-1}, using \disp{TS} with a different external potential term as in \disp{CalD} to write
\beq
&&\bra[H_J,X_i^{0  \si_i }(\tau)] X_f^{\si_f 0}(\tau') \ket=\half J_{ij}   \si_i \si \;   \bra X_i^{ 0 \si }(\tau)   X_j^{ \sib_i \sib }(\tau) X_f^{\si_f0}(\tau') \ket \nn \\
&=& - \half J_{ij} \; \left( \gamma_{ \sib_i \sib }(j, \tau)  - {\cal D}_{ \sib_i \sib }(j, \tau)   \right) \bra X_i^{ 0 \si }(\tau) X_f^{\si_f0}(\tau')\ket, \nn \\ \label{exchange-4}
\eeq
\sr{where the repeated spin index $\si$, and the site index $j$  are  summed over, while $\si_i,\si_f$ and site indices $i,f$ are held fixed.}
These two expressions \disp{exchange-3} and \disp{exchange-4} are alternate ways of writing the higher order Greens functions \cite{Engelsberg}. In order to describe a broken symmetry solution with superconductivity, we are required to use  \disp{exchange-3}, since using the other alternative  disconnects  the normal and anomalous Greens functions altogether, thereby precluding a superconducting solution.

The term $   \bra [\A_C(\tau),X_i^{0 \si_i}(\tau)]X_f^{\si_f 0}(\tau')\ket $ generates a term that is linear in  ${\cal J}$ which is treated similarly and the final result quoted in \disp{G0XY}.
 
We summarize these equations compactly by defining
\beq
G^{-1}_{0 i \si_i j \si_j}&=&\delta_{ij} \delta_{\si_i \si_j} \left(\chem - \partial_\tau \right) + t_{ij}  \delta_{\si_i \si_j} - \delta_{ij} {\cal V}_i^{\si_i \si_j}\nn \\
Y_{i \si_i j\si_j}&=&t_{ij} \gamma_{\si_i \si_j}(i, \tau) \nn \\
X_{i \si_i j\si_j}&=& - t_{ij} {\cal D}_{\si_i \si_j}(i, \tau) , 
\eeq
and write the exact equation
\beq
&& (G^{-1}_{0 i \si_i  j \si_j} - Y_{i \si_i j\si_j} - X_{i \si_i j\si_j}) \G_{j \si_j f \si_f}(\tau,\tau') = \delta(\tau-\tau') \delta_{if} (\delta_{\si_i \si_f} - \gamma_{\si_i \si_f}(i, \tau ))\nn \\
&&+ \half J_{ij}   \left( C_{ij}(\tau) - {\mathcal K}_{ij}(\tau) \right)\; \F_{j \si_i f \si_f}(\tau,\tau') \nn \\
&&+   \J_{j\si_j; i \si_k} \left( \delta_{\si_i, \si_k} - \gamma_{\si_i \si_k}(i, \tau )+ {\cal D}_{\si_i \si_k}(i, \tau )  \right) \si_j \F_{j \sib_j f \si_f}(\tau, \tau'), \label{G0XY}
\eeq
\sr{ where   the spins  $\si_j,\si_k$ and the site index $j$ are summed over, while $\si_i,\si_f$ and site indices $i,f$  are held fixed.
The final term drops off when we switch off the external potential ${\cal J}$.  Viewing the spin and site indices as joint matrix indices, these equations and their counterparts \disp{GOBACK},  are transformed into matrix equations below.}
\subsubsection{Greens function $\F$\label{EOM-F}}

The Gor'kov Greens function $\F$ in \disp{F}  satisfies  an exact  equation that can be found as follows. First we note
\beq
\partial_{\tau} \bra X_i^{ \sib_i 0}(\tau) X_f^{\si_f 0}(\tau') \ket = \bra[H_t+H_J+\A(\tau),X_i^{ \sib_i 0}(\tau)]\;\; X_f^{\si_f 0}(\tau')\ket
\eeq
A part of the right hand side satisfies
 \beq
   \bra[H_t+\A_{\rho} (\tau),X_i^{ \sib_i 0}(\tau)]\;\; X_f^{\si_f 0}(\tau')\ket =-\left( \chem \delta_{\si_i \si} - {\cal V}_i^{\sib_i \sib} \right)  \bra X_i^{ \sib 0}(\tau)\;\; X_f^{\si_f 0}(\tau')\ket \nn \\
   - t_{ij} \bra X_j^{ \sib_i 0}(\tau)\;\; X_f^{\si_f 0}(\tau')\ket + t_{ij} (\gamma_{\sib \sib_i}(i \tau)- {\cal D}_{\sib \sib_i}(i \tau))\bra X_j^{ \si 0}(\tau) ;\; X_f^{\si_f 0}(\tau')\ket,
  \eeq
  \sr{where the repeated spin index $\si$, and the site index $j$  are  summed over, while $\si_i,\si_f$ and site indices $i,f$ are held fixed.}
The exchange term is treated similarly to \disp{exchange-1}
  \beq
 [H_J,X_i^{  \sib_i 0 }]&=& \half J_{ij} \left(X_i^{  \uparrow 0} X_j^{ \downarrow 0 }- X_i^{ \downarrow 0 } X_j^{  \uparrow 0}\right) \si_i X_j^{0  \si_i }
  \eeq
so that using \disp{Kappastar} we get
\beq
\bra [H_J,X_i^{ \sib_i 0}]\; X_f^{\si_f 0}(\tau') \ket=  -\half J_{ij} \si_i  \left({C}^*_{i j }(\tau^-) - {\cal K}^*_{ i j}(\tau^-)   \right) \bra X_j^{0\si_i}(\tau) X_f^{\si_f 0}(\tau')\ket, \nn\\
\eeq
\sr{where the repeated internal index $j$ is summed over}

We gather and summarize these equation in terms of the variables  that  are ``time-reversed'' partners of \disp{G0XY} and hence denoted with hats:
\beq
\widehat{G}^{-1}_{0 i \si_i j \si_j}=\delta_{ij} \delta_{\si_i \si_j} \left(\chem +\partial_\tau \right) + t_{ij}  \delta_{\si_i \si_j} - \delta_{ij} {\cal V}_i^{\sib_i \sib_j}\nn \\
\widehat{Y}_{i \si_i j\si_j}=t_{ij} \gamma_{\sib_j \sib_i}(i, \tau)\nn \\
\widehat{X}_{i \si_i j\si_j}= - t_{ij} {\cal D}_{\sib_j \sib_i}(i, \tau)  \label{G0XYBack}
\eeq
So that 
\beq
\left( \widehat{G}^{-1}_{0 i \si_i j \si_j} - \widehat{Y}_{i\si_i j\si_j}- \widehat{X}_{i \si_i j \si_j} \right) {\cal F}_{j \si_j f \si_f}(\tau,\tau')= - \half J_{ij}   \left({C}^*_{ i j} - {\cal K}^*_{i j}   \right) \G_{j \si_i f \si_f }(\tau,\tau') \nn \\ 
+ \si_i \sum_{m }  \J^*_{ i \sib_n m\si_m}   (\delta_{\si_i, \si_n} -\gamma_{\sib_n \sib_i}(i, \tau)  + {\cal D}_{\sib_n \sib_i}(i, \tau)) \G_{m \si_m f \si_f}(\tau,\tau')\nn \\ \label{GOBACK}
\eeq
\sr{ where   the repeated spin indices    $\si_j,\si_n,\si_m$ and site index $j$ are summed over, while $\si_i,\si_f$  and $i,f$ are held fixed.}
The final term arising from $\bra [\A_C,X_i^{ \sib_i 0}] X_f^{\sib_f 0}(\tau') \ket $ drops off when we switch off the external potential ${\cal J}^*$.

\subsubsection{Summary of Equation in Symbolic Notation \label{Symbolic}}

 The equations  \disp{G0XY} and \disp{GOBACK}  are exact in the strong correlation limit.
 Noting that  all terms containing $\gamma$ and ${\cal D}$ in \disp{G0XY} and \disp{GOBACK} arise from  Gutzwiller projection,   we obtain the corresponding equations for the uncorrelated \tJ model in \disp{unctJ} by dropping these terms.  Recall also that the external potentials  ${\cal J,J}^*$ represent the imposed symmetry-breaking terms that force superconductivity, and are meant to be dropped at the end.  In this uncorrelated  case, let us understand the role of the terms with the Cooper pair derivatives ${\cal K,K}^*$.
 If we ignore  these terms   and also set  ${\cal J,J}^*\to 0$ right away, the equations \disp{G0XY} and \disp{GOBACK} reduce to the Gor'kov mean-field equations for the uncorrelated model \cite{Gorkov}, with the equation \disp{cstarf} providing a self consistent determination of $C_{ij}^*$ in terms of $\F$.
  Thus by neglecting the terms with ${\cal K, K}^*$, the role of the exchange $J$ is confined to providing the lowest order  electron-electron attraction   in the Cooper channel. This amounts to neglecting the  \sr{${\cal O}(J^2)$} dressings of the electron self energies and  irreducible interaction \sr{i.e. the pairing kernel in \disp{new-9}}. \sr{When retained, the normal state studies (see \refdisp{PRB-MaiShastry} Figs.~(22,23,24-(a))) show that the self energy terms arising  from $J$ change   the spectral functions  of the model only slightly. Regarding the irreducible interaction in the superconducting channel, the ${\cal O}(J)$ term is already attractive. Since we are in  the regime of $J\ll \mbox{max} \{ |t_{ij}|\}$  the retained term  is expected to dominate the neglected higher order term.   }
  \sr{In summary, strong Hubbard-Gutzwiller type short ranged  interactions renormalize} the Greens function to $\G$ from $G_0$, and the self energy terms due to $J$ are minor\cite{ECFL,PRB-MaiShastry}. The role of $J$ is significant \sr{ only  insofar as it provides}  a mechanism for superconducting pairing, \sr{ and potentially  magnetic instabilities  close to half filling. Keeping  these considerations} in mind,  we drop the terms involving ${\cal K, K}^*, {\cal J, J}^*$ in \disp{G0XY} and \disp{GOBACK}. This suffices for  our initial goal, of  generalizing    a Gor'kov type\cite{Gorkov}  mean-field treatment of \disp{unctJ}  to the   strongly correlated problem \disp{tJ}.

Multiplying the $\gamma$ and ${\cal D}$ terms, or equivalently the $X$ and $Y$ terms  with $\lambda$ and expanding the resulting equations systematically in this parameter constitutes the $\lambda$-expansion that we discuss below.

 With these remarks  in mind we make the following changes to the equations \disp{G0XY} and \disp{GOBACK}:
\begin{enumerate}
\item[(i)]  We  drop the terms proportional to ${\cal J, J}^*$ and the corresponding derivative terms ${\cal K, K}^*$. 
\item[(ii)]  Defining  the gap functions:
 \beq
\Delta_{ij}=  \half J_{ij} C_{ij} \;\;\mbox{and}\;\;
\Delta^*_{ij}=  \half J_{ij} {C}^*_{ij}  \label{gap-1}
 \eeq
 \item[(iii)]
 We scale the each occurrence of $\gamma,X,Y$,$ \widehat{X}, \widehat{Y}$ by $\lambda$.
\end{enumerate}

With these changes we   write the modified  \disp{G0XY} and \disp{GOBACK}:
  \beq
&&(G^{-1}_{0 i \si_i  j \si_j} - \lambda Y_{i \si_i j\si_j} - \lambda  X_{i \si_i j\si_j}) \G_{j \si_j f \si_f}\nn \\
&&= \delta(\tau-\tau') \delta_{if} (\delta_{\si_i \si_f} - \lambda  \gamma_{\si_i \si_f}(i, \tau ))
+   \Delta_{ij}\; \F_{j \si_i f \si_f} \label{For-1}
\eeq
\beq
\left( \widehat{G}^{-1}_{0 i \si_i j \si_j} - \lambda  \widehat{Y}_{i\si_i j\si_j}- \lambda \widehat{X}_{i \si_i j \si_j} \right) {\cal F}_{j \si_j f \si_f}=-    \Delta^*_{ij} \G_{j \si_i f \si_f } \label{For-2},
\eeq
\sr{where $\si_j$ is summed over in both \disp{For-1} and \disp{For-2}. Note that} the self consistency condition \disp{Cijstar} and \disp{cstarf} fix the correlation functions $C$'s  in terms of $\F$.
 As $\lambda \to0$ we get back  the meanfield equations of  Gor'kov for the uncorrelated-J model.   The $\lambda$ parameter governs the density of doubly occupied states, and hence a series expansion in this parameter builds in Gutzwiller type correlations systematically. We expand  the Greens functions to required  order in $\lambda$ and finally set $\lambda=1$.

    We write \disp{For-1} and \disp{For-2} symbolically as
\beq
(\GH_0^{-1} - \lambda Y - \lambda X). \G& =& (\bm{1} - \lambda \gamma) +  \Delta .\F  \label{For-3}\\
(\wh{\GH}_0^{-1} - \lambda \wh{Y} - \lambda \wh{X}). \F & =&  -  \Delta^* .\G \label{For-4}
\eeq
where the symbols $\G,\F$ etc are regarded as matrices in the space, spin and time variables, 
$\bm{1}$ is  the  Dirac  delta function in time and a Kronecker delta in space and spin, with the dot indicating matrix multiplication or time convolution. In the case of $X,\wh{X}$ it also indicates taking the necessary functional derivatives.

\section{Expansion of the Equations  in $\lambda$ \label{lambdaexpansion}}
We  decompose of both Greens functions in \disp{For-3} and \disp{For-4} as 
\beq
\G= \GH.\wmu, \;\; \F= \f. \wmu \label{factorize}
\eeq
where $\wmu$ is a function of spin, space and time that is  common to both Greens function.
As an example of the notation, the equation $\G=\GH.\wmu$ stands for
 $\G_{i \si_i j \si_j}(\tau_i,\tau_j)=\sum_{k \si_k}\int_0^\beta d\tau_k \;\GH_{i \si_i k \si_k}(\tau_i,\tau_k) \; \wmu_{k \si_k j \si_j}(\tau_k,\tau_j)  $. Here $\wmu$  is called the caparison (i.e. a further dressing) function, in a similar treatment of the normal state Greens function. The terms $\GH$ and $\f$ are called the auxiliary Greens function.
The basic idea   is that this type of factorization
can reduce \disp{For-3}, to a canonical type equation fo $\GH$, where the terms $\bm{1}- \lambda \gamma$ is replaced by $\bm{1}$.  We remark that this is a technically important step since the term  $\bm{1}- \lambda \gamma$  modifies the coefficient of the delta function in time, and encodes the distinction between canonical and non-canonical fermions.

To simplify further, we note that $X$ contains a functional derivative with respect to ${\cal V}$,  acting on objects to its right. When acting on a pair of objects, e.g. $X.\G=X.\GH.\wmu$, we generate two terms. One term is $ (X.g).\wmu$,
where the bracket, temporarily provided  here,  indicates that the operation of $X$ is confined to it. The second term has the derivative acting  on $\wmu$ only,  but the matrix product sequence is unchanged from the first term. We write the two terms together as
\beq
X.\GH.\wmu&=&\wick{\c X.\c \GH .\wmu}+ \wick{\c X.  \GH. \c \wmu}, \label{wicks-1}
\eeq
so that the `contraction' symbol refers to the  differentiation by $X$, and the `$.$' symbol refers to the matrix structure. We may view this as the  Leibnitz product rule.

Let us now operate with $X$ on the identity $\GH.\GHI= \iden$, where ${\bf g}^{-1}$ is the matrix inverse of $\GH$. Using  the Leibnitz product  rule, we find
\beq
\wick{ \c X. \c \GH}= - \left( \wick{ \c X .\bf{g} .\c {\bf g}}^{-1}\right).\GH \label{wicks-2}
\eeq
and hence we can rewrite \disp{wicks-1} in the useful form 
\beq
X.\GH.\wmu&=& - \left( \wick{ \c X .\bf{g} .\c {\bf g}}^{-1}\right).\GH
 + \wick{\c X.  \GH. \c \wmu} \label{wicks-3}.
\eeq
 
 With this preparation we rewrite \disp{For-3} the equation for $\G$  as
 \beq
 (\GH_0^{-1} - \lambda Y + \lambda \left( \wick{ \c X .\bf{g} .\c {\bf g}}^{-1}\right)). \GH .\wmu= (\bm{1} - \lambda \gamma) +   \Delta.\f.\wmu +  \lambda  \wick{\c X.\GH.\c \wmu} \label{For-5}
 \eeq
  We now choose
 $\GH,\f$ such that
  \beq
 (\GH_0^{-1} - \lambda Y + \lambda \left( \wick{ \c X .\bf{g} .\c {\bf g}}^{-1}\right)). \GH = \iden +  \Delta. \f . \label{new-1}
 \eeq
Substituting \disp{new-1} into \disp{For-5},  we find that     $\wmu$  satisfies the equation
 \beq
 \wmu= (\bm{1} - \lambda \gamma) +  \lambda \wick{\c X.\GH.\c \wmu}. \label{new-2}
 \eeq
 Note that \disp{new-1} has the structure of a canonical equation since we replaced the $\bm{1}-\lambda \gamma$ term by $\bm{1}$ in \disp{For-5}. 
 Thus the non-canonical \disp{For-3}  for $\G,\F$ is replaced by  a pair of canonical equations for $\GH,\wmu$. In \disp{new-1} we note that the action of $X$ is confined to the bracket $\lambda \left( \wick{ \c X .\bf{g} .\c {\bf g}}^{-1}\right)$, unlike the term $\lambda X.\G$ in the  initial \disp{For-3} . We may thus view the term in bracket in \disp{new-1} as a proper self energy for $\GH$.

 For treating the equation for $\F$ \disp{For-4} we use the same scheme \disp{factorize} and find 
\beq
\wh{X}.\F = \wh{X}.\f.\wmu 
= -\left( \wick{\c{\wh{X}} .  \f  . \c \f^{-1}} \right). \f . \wmu + \wick{\c{\wh{X}} .\f. \c\wmu}
\eeq
With this we rewrite \disp{For-4} after   cancelling an overall right multiplying factor  $\wmu$
   \beq
 (\wh{\GH}_0^{-1} - \lambda\wh{Y} + \lambda  \wick{\c {\wh{X}}. \f . \c \f^{-1} } ). \f  & =&   - \Delta^* .\GH + \lambda  \wick{\c{\wh{X}} .\f. \c\wmu}. \wmu^{-1} \label{new-3}
 \eeq 
 Summarizing we need to solve for $\f,{\bf g}, \wmu,\Delta^*$  from Eqs.~(\ref{new-1},\ref{new-2},\ref{new-3}) by iteration in powers of $\lambda$. 
 
 \subsection{Simplified  Equations  near  $T_c$\label{nearTc}}

 For the present work, we note that the equation \disp{new-3} simplifies considerably, if we work close to $T_c$. In this regime $\f$ may be assumed to be very small, enabling us to throw away all terms of ${\cal O}(f^2)$ and also to discard terms of ${\cal O}( \lambda f)$.  This  truncation  scheme  is sufficient to determine $T_c$ for low orders in $\lambda$.
 
 When $T\sim T_c$,  throwing away terms of ${\cal O}(f^2)$ and ${\cal O}( \lambda f)$, we obtain the simplified version of \disp{new-3} 
 \beq
 \f= -\wh{\GH}_0.\Delta^* .\GH + o(\lambda \f), \label{new-4}
 \eeq
so that \disp{new-1} can be written as
\beq
\GH^{-1}=\GH_0^{-1} - \lambda Y + \lambda \left( \wick{ \c X .\bf{g} .\c {\bf g}}^{-1}\right)+\Delta.\wh{\GH}_0.\Delta^*   \label{new-5}
\eeq
In this limit  the above two   are the  ${\cal O}(\lambda^2)$ equations  required to be solved, together with \disp{new-2}  and the self consistency condition \disp{gap-1}, \disp{cstarf}. The latter   can be combined with \disp{F} as
\beq
\Delta_{ij}^*&=& \half J_{ij} {C}^*_{ij}=- \half J_{ij} \sum_\si \F_{i \si, j \si}(\tau^+, \tau) \label{new-6}
\eeq
and further reduced using \disp{factorize}.
On turning off the external potentials we regain time translation invariance.  We next perform a fourier transform to fermionic Matsubara frequencies $\omega_n = \frac{\pi}{\beta} (2 n+1)$ using the definition $\F(\tau)= \frac{1}{\beta}\sum_n e^{- i \omega_n \tau} \F(i \omega_n)$, and write \disp{factorize} in the frequency domain   as 
\beq
\F_{p \si}(i \omega_n) = \f_{p \si}(i \omega_n) \wmu_{p \si}( i\omega_n). 
\eeq
Thus taking spatial fourier transforms with the definition 
\beq
J(q) = 2 J \left( \cos q_x+ \cos q_y\right),
\eeq
so that  the self consistency condition   \disp{new-6} finally reduces to
\beq
\Delta^*(k) & =& - \frac{1}{2 \beta}\sum_{p \si \omega_n} J(k-p) \f_{p \si}(i \omega_n) \wmu_{p \si}(i \omega_n) \label{new-7}
\eeq
We may write \disp{new-4} as
\beq
\f_{p \si}(i \omega_n) = - \wh{\GH}_{0 \si}(p, i \omega_n) \Delta^*(p) \GH_{\si}(p, i \omega_n)
\eeq
where the time reversed free Greens function
\beq
\wh{\GH}_{0 }(p, i \omega_n)= \frac{1}{- i \omega_n +\chem_0 - \varepsilon_{-p}}= \frac{1}{- i \omega_n - \xi_{p}}
\eeq
with $\xi= \varepsilon_p- \chem_0$ and by
using  $\varepsilon_p=\varepsilon_{-p}$, and $\chem_0$ is taken as the non-interacting system chemical potential, discarding  the corrections of $\mu$  due to $\lambda$.
Therefore \disp{new-7} becomes
\beq
\Delta^*(k) &=&  \frac{1}{2\beta}\sum_{p \si \omega_n} J(k-p) \wh{\GH}_{0 \si}(p, i \omega_n) \Delta^*(p) \GH_{\si}(p, i \omega_n) \wmu_{p \si}(i \omega_n) \nn \\ \label{new-8}
\eeq
Here $\GH$ is taken from \disp{new-5}, i.e. the ${\cal O}(\lambda^2)$ Greens function with a small correction (for $T\sim T_c$)   from the gap $\Delta$.
Performing  the spin summation and recombining $\GH .\wmu=\G$,  we get the equation in terms of the  physical electron Greens function
\beq
\Delta^*(k) &=&   \frac{1}{\beta}\sum_{p  \omega_n} J(k-p) \Delta^*(p) \wh{\GH}_{0 }(p, i \omega_n)  \G(p, i \omega_n). \label{new-9}
\eeq
This is an important result of our formalism, it represents the leading order Gor`kov equation for the \tJ model. It is analogous to a refinement of Gor`kov's equation  \cite{Gorkov}, usually called the Eliashberg equation \cite{Eliashberg}, valid for  strong electron-phonon coupling superconductivity. Our $\lambda$ expansion  plays the role of the Migdal theorem \cite{Migdal} in that problem.  The analogy  with Migdal \cite{Migdal} and Eliashberg's \cite{Eliashberg}  work is only superficial,
 since the strongly correlated problem  does not share the  physics of  the separation of the electronic and phonon  time scales, underlying those results. 
 
In \disp{new-9} the physical electron Greens function $\G$ is taken from  the ${\cal O}(\lambda^2)$ theory if we neglect the corrections from the gap, which vanishes above $T_c$ anyway.
We  express the physical Greens function in terms of its spectral function $A(p,\nu)$
\beq
\G(p, i \omega_n)= \int d \nu \; \frac{ A(p,\nu)}{i \omega_n - \nu} \label{spectral-rep}
\eeq
 The frequency integral in \disp{new-6} can be performed as 
\beq
\frac{1}{\beta}\sum_{  \omega_n}  \wh{\GH}_{0 }(p, i \omega_n)  \G(p, i \omega_n)=   \int d \nu \;  A(p,\nu) \frac{1-f(\nu)-f(\xi_p)}{\nu+\xi_p}.
\eeq
where $f$ is the fermi distribution $f(\nu)=1/(1+ \exp{\beta \nu})$.
Hence
\beq
\Delta^*(k)= \sum_p J(k-p) \Delta^*(p) \;  \int d \nu \;  A(p,\nu) \frac{1-f(\nu)-f(\xi_p)}{\nu+\xi_p}. \label{imp-1}
\eeq
In summary this  eigenvalue type equation for $\Delta^*(k)$, together with the spectral function $A(p,\nu)$ determined from the  ${\cal O}(\lambda^2)$ Greens function in \disp{new-5},  gives the self-consistent gap near $T_c$. At sufficiently high temperatures, i.e. in the normal state $T>T_c$  $\Delta^*$ vanishes, so that $A$ is independent of $\Delta^*$. In this case  \disp{imp-1} reduces to a linear integral equation for $\Delta^*$.
 We may then determine $T_c$ from  the condition that the largest eigenvalue crosses 1.  For this purpose we only  need  the normal state electron  spectral function of the strongly correlated metal.

\section{Estimate of $T_c$\label{EquationforTc}}
\subsection{Equation for determining  $T_c$}
The condition for obtaining a d-wave superconducting state is given by setting $T=T_c^+$ in
\disp{imp-1} 
writing $\Delta^*(k)= \Delta_0(\cos k_x-\cos k_y)$, using the normal state spectral function for $A$ and canceling an overall factor $\Delta_0 (\cos k_x-\cos k_y)$. Following these steps   we get
\beq
1= J \sum_{p}\left\{\cos(p_x)-\cos(p_y) \right\}^2  \;  \int d \nu \;  \frac{1-f(\nu)-f(\xip)}{\nu+\xip}
 A(p,\nu)\bigg|_{T_c}. \label{Tc1}
\eeq
Instead of working with \disp{Tc1}, it is convenient to make a useful simplification for the average over angles.
Since  \disp{Tc1} is largest when   $\vec{p}$ is  on the fermi surface, we  factorize the two terms  and write
\beq
1&=& J  \Psi(\mu_0) \; \Gamma \label{Tc20} \\
\Gamma&=&\sum_{p}  \int d \nu \;  \frac{1-f(\nu)-f(\xip)}{\nu+\xip}
 A(p,\nu)\bigg|_{T_c} \label{Tc2}
 \eeq
 where $\Gamma$ is a particle-particle type susceptibility.  Here $\Psi(\mu_0)$ is more correctly the weighted average of
 $\left\{\cos(p_x)-\cos(p_y) \right\}^2$ with a  weight function  that is the integrand in \disp{Tc2}. We simplify it to the fermi surface averaged momentum space d-wavefunction
 \beq
\Psi(\mu_0)&=&  \frac{1}{n(\mu_0) }\sum_{p}\left\{\cos(p_x)-\cos(p_y) \right\}^2 ) \delta(\varepsilon_p-\mu_0) \label{D-wave}
\eeq
where  $n(\epsilon)$ is the band density of states (DOS) per spin and per site, at energy $\epsilon$,
\beq 
n(\epsilon) &=&\frac{1}{N_s} \sum_{p} \delta(\varepsilon_p-\epsilon).
\eeq
Using this simplification and performing the angular averaging over the energy surface $\varepsilon_{\vec{p}}=\epsilon$ we write the (particle-particle) susceptibility $\Gamma$ (\disp{Tc2})  as
\beq
\Gamma=  \int \; d\epsilon \int d \nu \;  n(\epsilon) \; A(\epsilon,\nu) \; \frac{1-f(\nu)-f(\epsilon-\mu_0)}{\nu+\epsilon-\mu_0}\bigg|_{T_c}
 . \label{Tc3}
\eeq
where $A(\epsilon,\nu)$ is the angle-averaged version of the spectral function $A(p,\nu)$. We estimate this expression below for the extremely correlated fermi liquid, by using a simple model for the spectral function $A$.

In  \disp{Tc3} if we replace  the spectral function $A$ by  the (fermi gas) non-interacting  result $A_{0}(\epsilon,\mu_0)=\delta(\nu-\epsilon+\mu_0)$, we obtain the Gorkov-BCS mean-field theory, where the susceptibility $\Gamma$ reduces to   $ \int d\epsilon  \;  n(\epsilon)  \; \frac{\tanh\half \beta_c (\epsilon-\mu_0)}{2(\epsilon-\mu_0)} $. This is
 evaluated  by expanding  around the fermi energy, and  utilizing the  low T formula  $\int _0^{W_0} \frac{d \epsilon}{\epsilon} \, {\tanh\half \beta_c \epsilon}\sim \log\left[ \frac{\zeta_0 W_0}{k_B T_c}\right] $, where $W_0 $ is  the half-bandwidth and
 $\zeta_0= 1.13387\ldots$. Equating $\Gamma$  to $1/J \ppsi$ gives the  d-wave superconducting transition temperature for the uncorrelated  \tJ model  
 \beq
k_B T_c^{(un)}&\sim&   1.134\; W_0 \; e^{-\frac{1}{g}},\label{Tcuncorr}
\eeq
with the  superconducting  coupling constant
\beq
g&=&  J \Psi(\mu_0) n(\mu_0) \label{g}.
\eeq

\subsection{Model spectral function \label{modelspectral}}
 We   next use a simple model spectral function to estimate these integrals. It has the great advantage that we can carry out most integrations analytically and get approximate but  closed form analytical expressions for $T_c$, which provide useful insights.  The model spectral function contains the following essential features of strong correlations namely:
  \begin{itemize}
 \item   A quasiparticle part with  fermi liquid type parameters, where the quasiparticle weight $Z$ goes to 0 at half filling $n=1$, and 
 \item  A   wide    background. 
 \end{itemize}
 The  model  spectral function used is
 in the spirit of Landau's fermi liquid theory\cite{Landau,Nozieres,AGD} 
 with suitable modifications due to strong correlation effects\cite{ECFL}.
 We take the spectral function as 
 \beq
A(\epsilon, \nu) = Z \delta( \nu - \mbstar \epsilon) + (1-Z) \frac{1}{2 W_0} \Theta(W_0- |\nu|).
\label{FLT} \eeq
 Here $\Theta(x)=\frac{1}{2}(1+\frac{x}{|x|})$,  $W_0$  the half-bandwidth 
  $\mbstar$ is the renormalized effective mass of the fermions,  and $Z$  is the  fermi liquid renormalization factor.  The first term is the  quasiparticle part with weight $Z$,  and  second part represents the background modeled as  an inverted square-well. 
Integration over $\nu$ gives unity at each energy $\epsilon$.  $Z$ is  chosen to reflect the fact that we are dealing with a doped Mott-Hubbard insulator  so it must vanish at $n=1$. For providing a simple estimate we use    Gutzwiller's result \cite{Gutzwiller,Brinkman}
\beq
Z=1-n. \label{GutzZ}
\eeq
 The  effective mass is related to $Z$ and the k-dependent Dyson self energy $\Sigma$ through the standard fermi liquid theory\cite{Landau,Nozieres,AGD} formula
\beq
\mbstar= Z \times (1+ \frac{\partial \Sigma(\vec{k}, \mu)}{\partial \varepsilon_k}\bigg|_{k_F}) \label{mstarFL}.
\eeq
The Landau fermi liquid renormalization factor $\mbstar$ can  be inferred from heat capacity experiments  provided   the bare density of states is assumed known. 

Using \disp{FLT} in \disp{Tc3} and decomposing the  susceptibility $\Gamma$
into a quasiparticle and background part, the equation determining $T_c$ is:
\beq
\left( \Gamma_{QP}+\Gamma_{B}\right)\bigg|_{T\to T_c} & = &\frac{1}{J \ppsi } \label{Gorkov1} \\
\Gamma_{QP}&=& Z \int d\epsilon \, n(\epsilon) \frac{1-f(\epsilon-\mu_0)-f(\mbstar (\epsilon-\mu_0))}{(\epsilon-\mu_0) (1+\mbstar)} \\
\Gamma_{B}&=&\frac{(1-Z)}{2 W_0}\int d\epsilon \, n(\epsilon)\int_{-W_0}^{W_0} d\nu   \frac{1-f(\epsilon-\mu_0)-f(\nu)}{(\epsilon-\mu_0)+\nu }.\nn \\ 
\eeq

Using the same approximations that lead to \disp{Tcuncorr} the $\Gamma_{QP}$ can be evaluated as
\beq
\Gamma_{QP}&=& \frac{Z n(\mu_0)}{1+\mbstar}  \int_0^{W_0}  \frac{d\epsilon}{\epsilon} \ \left( \tanh  \frac{\epsilon}{2 k_B T}- \tanh   \frac{\epsilon \, m/m^*}{2 k_B T} \right),
\eeq
and hence at low enough  $T$ the estimate
\beq
\Gamma_{QP}\sim n(\mu_0) \frac{2 Z}{1+\mbstar} \log\left[\frac{\zeta_0 W_0 \sqrt{\mbstar}}{k_B T} \right]. \label{GammaQP}
\eeq 
Unlike the quasiparticle part with this  $\log T$ behavior at low T,
the background part  is nonsingular as $T\to0$,   since   a double integral over the region of small $\epsilon-\mu_0$ and $\nu$ is involved. It can be estimated by setting $T=0$,  $\epsilon-\mu_0\sim\epsilon$ and replacing $n(\epsilon)\sim n(\mu_0)$. With 
\beq
\Gamma_B&\equiv& n(\mu_0) \gamma_B, \\
 \gamma_B&=&\frac{(1-Z)}{2 W_0}   \int_{-W_0}^{W_0}\int_{-W_0}^{W_0} \half \frac{\sign(\epsilon)+\sign(\nu)}{\epsilon+\nu}\,d \epsilon \, d\nu.
\eeq
Integrating this expression we obtain 
\beq
\gamma_B=(1-Z)  \log 4. \label{gamma}
\eeq

Combining Eqs.~(\ref{Gorkov1},\ref{GammaQP},\ref{gamma})  we find
\beq
k_BT_c &\sim& 1.134 \; W_0 \times  \sqrt{ \mbstar} \times e^{-\frac{1}{g_{eff}}} \label{Tc}
\eeq
where the effective superconducting coupling: 
\beq
g_{eff}&=& \frac{ 2 Z}{\left(1+\mbstar\right)} \left\{J_{eff} \Psi(\mu_0) n(\mu_0)\right\} \label{geff}
\eeq
and an effective exchange 
 \beq
 J_{eff}= \frac{J}{1- \gamma_B J \ppsi n(\mu_0)}, \label{Jeff}
 \eeq
where the denominator represents an  enhancement due to  the  background spectral weight.
 In comparing \disp{Tc} with the uncorrelated result \disp{Tcuncorr}  several  changes are visible.
  The bandwidth prefactor  is reduced by correlations due to the factor of $\sqrt\mbstar\ll 1$. This factor vanishes as $n\to 1$ thereby diminishing superconducting $T_c$ in the close proximity of the  insulator. A similar but even more drastic effect arises from multiplying factor $\frac{2 Z}{ (1+\mbstar)}$ in the coupling $g_{eff}$ \disp{geff}.
This term reflects the  quasiparticle  weight in the pairing process, and  vanishes  near  the insulating state.  Being situated in the exponential, it  kills superconductivity even more effectively than the bandwidth prefactor. Away from the close proximity of the insulator  other terms in $g_{eff}$ become prominent, allowing for the possibility of superconductivity.
Amongst them  is the replacement of the exchange energy  by  $J_{eff}$. In a density range where $\ppsi n(\mu_0)$ is appreciable, this enhances $J_{eff}$ over $J$ due to the feedback nature of \disp{Jeff},  and has an important impact on determining  the   phase region with superconductivity.

 
 \subsection{Numerical Estimates of  $T_c$\label{NumericalestimatesofTc}}
We turn to the task of   estimating the order of magnitude of the Tc in this model.  When we take typical values for cuprate systems:  $W_0\sim 10^4$K (i.e $\sim$1 eV) and $J\sim 10^3$K (i.e. $\sim$0.1 eV),
the  transition temperature of  the uncorrelated model $T_c^{(un)}$ \disp{Tcuncorr} is a few  thousand K, at most   densities. It remains robustly non-zero at half filling, since in this formula correlation effects are yet to be built in and the Mott-Hubbard insulator is missing. For the correlated system,
we  estimate $T_c$ from \disp{Tc} using  similar values of model parameters.  The terms arising from correlations in \disp{Tc} are guaranteed to suppress superconductivity near the insulating state, since $Z\to0$ and the quasiparticle is lost. A more refined   question  is    whether  an   intermediate density regime  ($ \delta > 0$)  can support superconductivity. And  if  so,   whether the temperature  scales are  robust enough to be observable. Within the context and confines of the simplified model spectral function considered, we answer both questions positively here.

\subsubsection{Choice of model parameters \label{ModelParameters}} 

In order to estimate the order of magnitude of the Tc,  its  dependence on $J$ and band  parameters, we choose parameters similar to those used in contemporary studies for the single layer High $T_c$ compound $La_{2-x}Sr_xCuO_4$.  The hopping Hamiltonian   
$-\sum_{ij} t_{ij} \widetilde{C}_{i \sigma}^\dagger \widetilde{C}_{j \sigma}$, gives rise to band energy dispersion  $\varepsilon(\vec{k})= -2 t (\cos k_x +\cos k_y ) - 4 t' \cos k_x \cos k_y- 2 t'' (\cos 2 k_x + \cos 2 k_y)$ on a square lattice. Thus the hopping amplitudes $t_{ij}$  are  equal to $t$ when $i,j$ are nearest neighbors,  $t'$ when $i,j$ are second-nearest neighbors, and $t''$ when $i,j$ are third-nearest neighbors.  For this system we will use the values \cite{ARPES-LSCO,Ogata}
\beq
t=0.45 \, eV, \;\; t'/t = - 0.16 \pm 0.02, \;\; t''/t=.01. \label{Hoppings}
\eeq
This parameter set is roughly consistent with the experimentally determined fermi surface of $La_{2-x}Sr_xCuO_4$ \cite{ARPES-LSCO}, we comment below on considerations leading to a more precise choice.
 The tight binding band extends from $-W_{0}\leq \epsilon \leq W_0$, where $W_0=  4 t$, neglecting a small shift due to $t'$. 
The exchange energy is chosen to be
\beq
J/t =0.3, \;\; \mbox{or} \;\;J/k_B\sim 1550K, 
\eeq
as determined from two magnon Raman experiments\cite{Rajiv} on the parent insulating $La_2CuO_4$. Note  that the \tJ model is  obtainable from the Hubbard model by performing a large $U/t$  super-exchange expansion, giving  $J= \frac{ 4 t^2}{U}$. Thus our  choice of $J$ corresponds to  a strong coupling type magnitude of $U/t\sim 13.3$ in the Hubbard model, placing it in a perturbatively inaccessible regime   of that model.

We  now discuss the enhancement of effective mass $\mbstar$ \cite{Loram}. 
In the proximity of the Mott-Hubbard  insulating state $n\to 1$, an enhancement in $\frac{m^*}{m}$ is expected on general grounds, reflecting a diminished thermal excitation energy scale due to band narrowing.  For illustrating the role of this parameter we use two complementary estimates
\beq
\mbstar &\sim& 3.4 \; (1-n),  \;\;\mbox{(a)} \label{mstarA} \\
\mbstar &=& Z, \;\;\;\;\;\;\;\;\;\;\;\;\;\;\;\;\mbox{(b)} \label{mstarB} 
\eeq
where  estimate (a) gives a two-fold enhancement of  $\frac{m^\star}{m}$ at $\delta=.15$,  while  estimate (b), obtained by neglecting the $k$ dependence of the self energy in \disp{mstarFL},
gives a seven-fold enhancement. The formulas used are simple enough so that the effect other estimates for $\mbstar$ should be easy enough for the reader  to gauge.

\subsubsection{Results  \label{Results}}

\begin{figure}[htbp]
\centering
\includegraphics[width=.72\columnwidth]{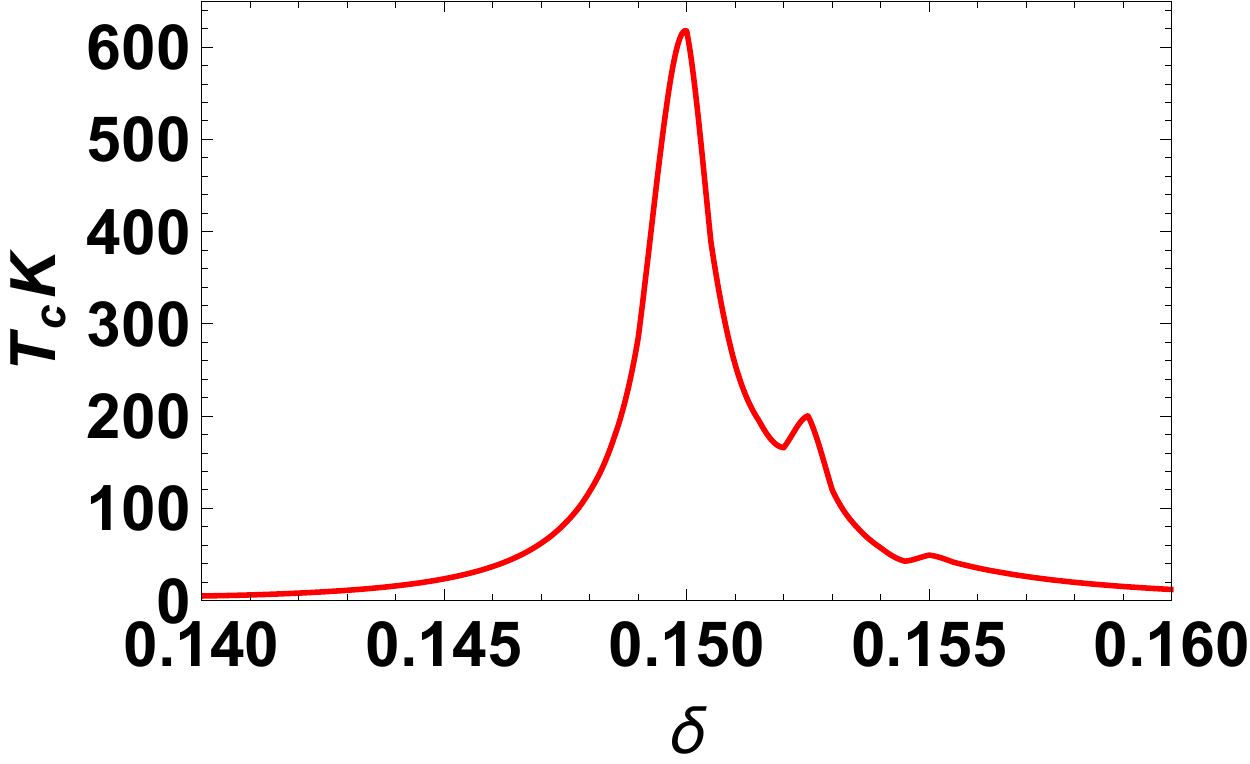}
\caption{\footnotesize 
  The superconducting transition temperature for the correlated model $T_c$ (\disp{Tc})  \label{Fig-1-Tc} ($t'/t=-0.159, \; t''/t=0.01, \; \frac{m}{m^*}=Z$). 
  The scale of the maximum transition temperature is smaller by an order of magnitude from  the uncorrelated model. As  the insulator is approached $\delta \to 0$, and $T_c$  decreases drastically. This is easy to understand  since the quasiparticle weight
$Z$ shrinks on approaching the insulating state, killing the coupling $g_{eff}$ \disp{geff}. 
When $\delta$ goes beyond the peak (optimum) value, the effective superconducting coupling $g_{eff}$ agains falls off as seen in \figdisp{Fig-symmetry} and in \figdisp{Fig-geff} due to the other factors in \disp{geff}. When   $g_{eff}$ drops  below $\sim0.12$,    the resulting  $T_c$  is negligible.
 }
  \end{figure}

\begin{figure}[htbp]
\centering
\includegraphics[width=0.72\columnwidth]{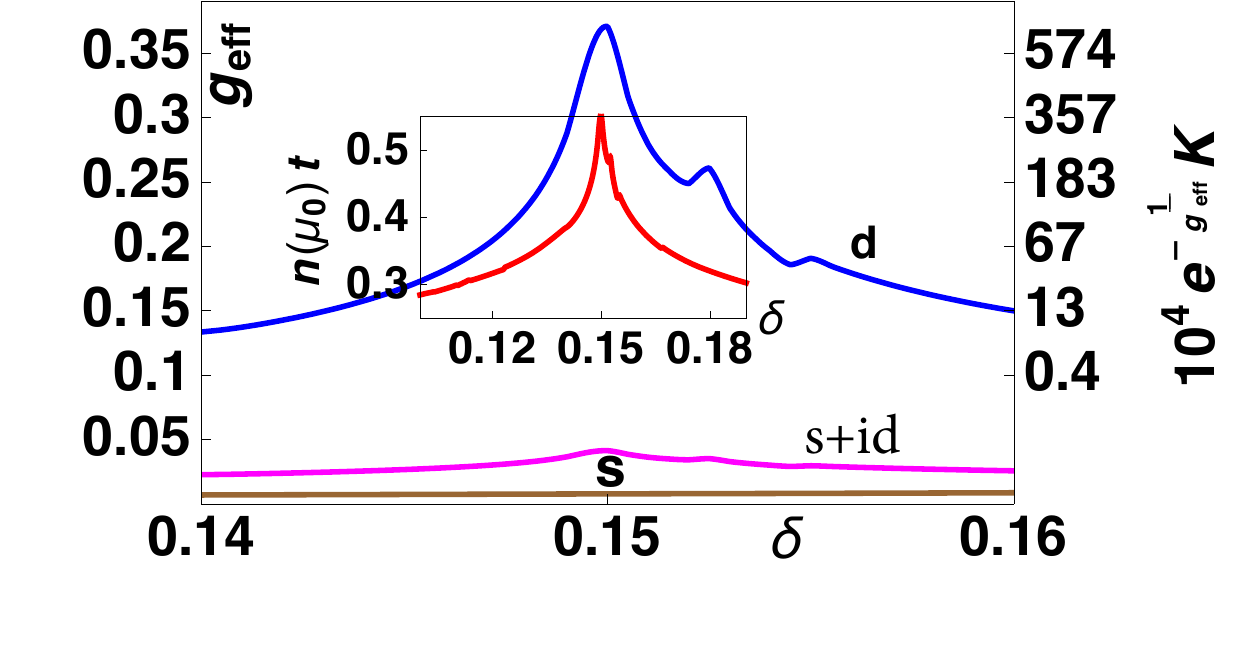}
\caption{\footnotesize The  figure and inset use   $t'/t=-0.159$, $t''/t=0.01$, and $\frac{m}{m^*}=Z$.
The effective superconducting coupling $g_{eff}$ (\disp{geff})  for  three Cooper pair symmetries: (i)(blue) d-wavefunction  $\langle \left\{\cos(k_x)-\cos(k_y)\right\}^2 \rangle_{FS}$, (ii)(brown) extended s-wavefunction    $\langle \left\{\cos(k_x)+\cos(k_y)\right\}^2 \rangle_{FS}$, and (iii)(magenta)   $s+i d$ wavefunction  $\langle \left\{\cos^2(k_x)+\cos^2(k_y)\right\} \rangle_{FS}$.  For the d-wavefunction, the drastic decrease of $T_c$ on both sides of the peak values in 
\figdisp{Fig-1-Tc}  can be understood by referring to the the second y-axis, giving the temperature scale $T_c^{appx}= 10^4 \times e^{- \frac{1}{g_{eff}}}$ K. This scale  provides an  order of magnitude of  $T_c$ at  a given $g_{eff}$ by assuming a prefactor $10^4$K. It illustrates the rapid reduction of $T_c$   when $g_{eff}\lessim0.12$. The other two symmetries  lead to much smaller couplings and are therefore  ineffective. {\bf Inset:} The band DOS at the fermi energy shows  an enhancement around  the hole density $\delta=0.15$.
 \label{Fig-symmetry}  }
 \end{figure}

 In \figdisp{Fig-1-Tc} the superconducting transition temperature for d-wave symmetry is shown as a function of the hole density $\delta=1-n$ where the band parameters are indicated in the caption. It shows that $T_c$ is maximum at $\delta\sim0.15$ and falls off rapidly as one moves away from that density in either direction. The scale of $T_c$ is a few hundred K, which is an order of magnitude lower than that of the uncorrelated system. The small  kink-like features to the right of the peak reflect structure in the DOS shown as inset in \figdisp{Fig-symmetry}.
 In \figdisp{Fig-symmetry}  the effective superconducting coupling $g_{eff}$ is shown for three different symmetries of the Cooper pairs: $d$-wave, extended $s$-wave, and $s+i d$-wave. It is clear that within this theory, only  d-wave symmetry leads to robust superconductivity,  the other two symmetries lead to    effects too small  to be observable. 
 \begin{figure}[htbp]
\centering
\includegraphics[width=.72\columnwidth]{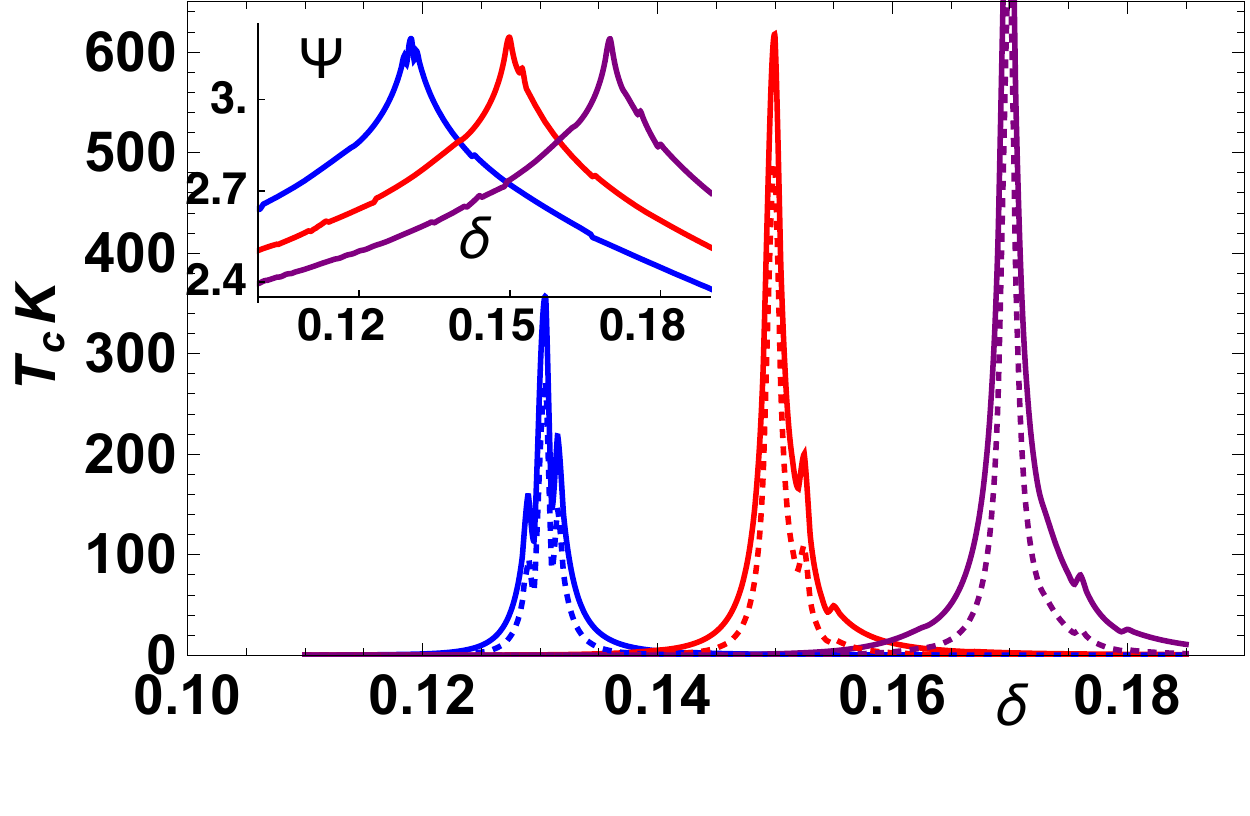}
\caption{\footnotesize 
  The superconducting transition temperature for the correlated model $T_c$ (\disp{Tc})  \label{Figure-Tc-Second}  for  three parameter sets- (i) (red) $t'/t$$=-0.159$, $t''/t$$=0.01$ with $\delta_{peak}=0.15$,  (ii) (blue) $t'/t$$=-0.137$, $t''/t=.01$  with $\delta_{peak}=0.13$ and (iii)(purple)
$t'/t$$=-0.181$,$t''/t$$=0.01$ with $\delta_{peak}=0.17$. The solid lines use $\frac{m}{m^*}=Z$ for  and the dashed lines
  $\frac{m}{m^*}=3.4 \delta$.
  {\bf Inset:}
 $\Psi(\mu_0)$  the fermi surface averaged d-wavefunction $\langle\left\{\cos(k_x)-\cos(k_y)\right\}^2\rangle_{FS}$  is shown for the three sets of band parameters. The peak values occur at the densities where $T_c$ is highest.  Their peak magnitude $\sim 3.2$ indicates a  strong constructive interference effect from   $\vec{k}\sim\{ \pm \pi,0\},\{ 0,\pm \pi\}$, where $|\cos(k_x)-\cos(k_y)|\sim 2$.   }
  \end{figure}
  From \figdisp{Figure-Tc-Second} we see that the peak density  is shifted by varying the band hopping parameters. As the peak density moves towards small $\delta$, its height  falls rapidly. This is understandable as the effect of the quasiparticle weight  $Z$ in the formulas Eqs.~(\ref{Tc},\ref{geff}). We also note that the use of different expressions for the effective mass  in Eqs.~(\ref{mstarA},\ref{mstarB}) change the width of the allowed regions somewhat, but are quite comparable. 

  The inset in  \figdisp{Figure-Tc-Second} displays the d-wavefunction averages corresponding to   the same   sets of parameters. It is interesting to note that the height of the peaks  $\Psi_{max}$$\sim$$3.2$ are close to their upper bound $4$, from a type  of  constructive interference that requires comment. Note first that  the DOS can be expressed as a line integral in the octant of the Brillouin zone  $n(\mu_0)=  \frac{2}{\pi^2} \int_0^{k_{max}} \frac{d k_x}{|v^y{(\vec{k})}|}$,
 where the velocity $v^y{(\vec{k})}= 2  \sin(k_y) (t+ 2t' \cos k_x + 4 t'' \cos k_y)$ is evaluated with $k_y\to k_y(k_x,\mu_0)$ on the fermi surface.
 Thus the region of small $|v^y|$ dominates the integral. If $v^y$ vanishes  on the fermi surface, we get a (logarithmic van Hove) peak in the DOS. Now the average of $\Psi(\mu_0)$ 
is largest, when $\vec{k}$ is close to $\{\pm \pi,0\}$ and $\{0, \pm \pi\}$. Therefore if   the fermi surface passes through $\{\pm \pi,0\}$ and $\{0, \pm \pi\}$ for an ``ideal density'', then we simultaneously maximize the average of $\Psi$, {\em and} obtain a large $DOS$.
The condition for this is found by equating the band energy at $\{\pm \pi,0\}$ to the chemical potential $\mu_0= 4 t'-4 t''$, thereby fixing the corresponding density $\delta$. It follows that a given  $\delta$ can be found from several different  sets of the parameters $t', t''$.  The inset of \figdisp{Figure-Tc-Second} shows the average $\ppsi$ displays peaks, the middle one (red)  coincides in location with the peak in the DOS in the inset of \figdisp{Fig-symmetry}.
\begin{figure}[htbp]
\centering
\includegraphics[width=.63\columnwidth]{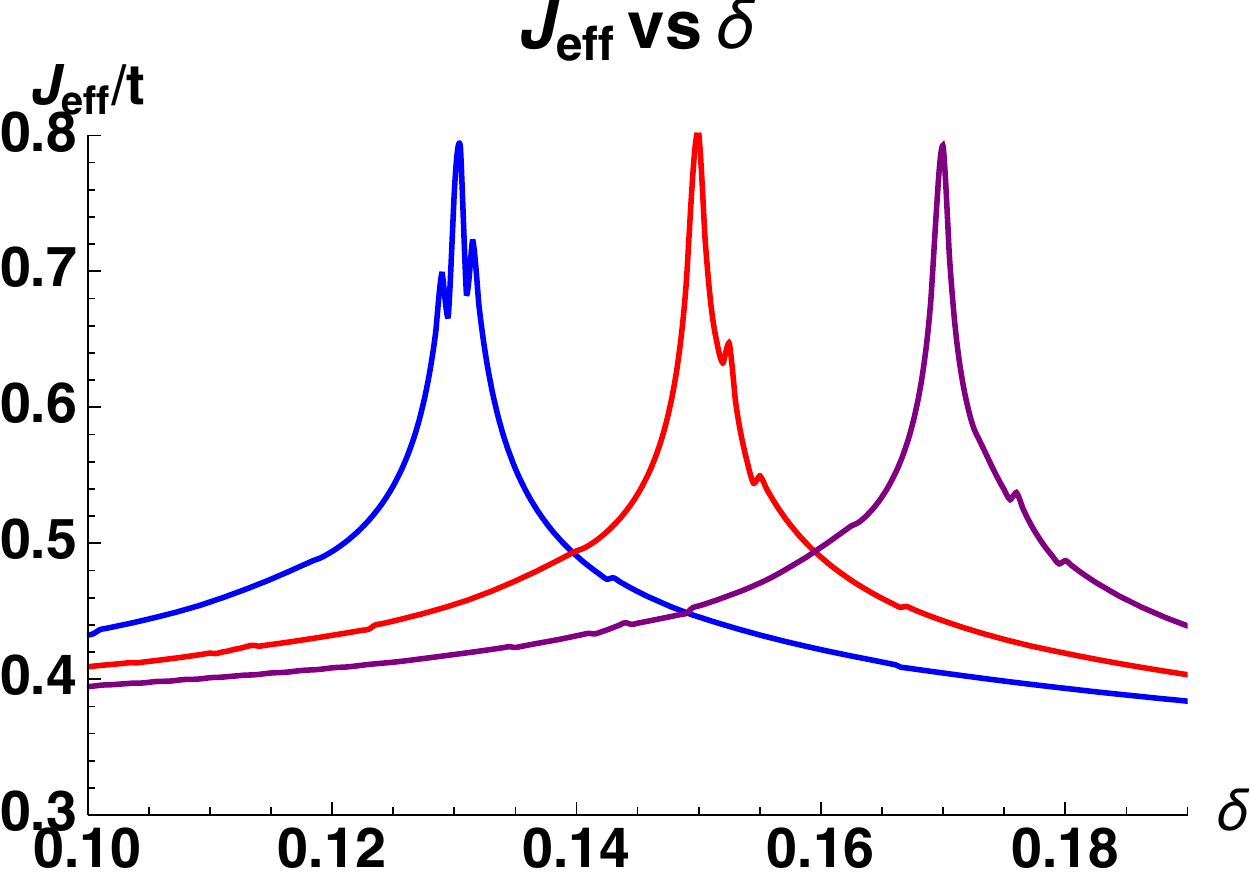}
 \caption{\footnotesize
 The effective exchange $J_{eff}$  from \disp{Jeff}  for the  
  three parameter sets- (i) (red) $t'/t$$=\!-0.159$, $t''/t$$=0.01$ with $\delta_{peak}=0.15$,  (ii) (blue) $t'/t$$=\!-0.137$, $t''/t=.01$  with $\delta_{peak}$$=$$0.13$ and (iii)(purple)
$t'/t$$=\!-0.181$, $t''/t$$=0.01$ with $\delta_{peak}$$=$$0.17$.
    Since we assumed $J/t\sim 0.3$  (\disp{Jeff}),    $J_{eff}/t$ is   considerably enhanced in  the  range of densities exhibiting high $T_c$.  This enhancement in turn boosts up $g_{eff}$, via  \disp{geff}, and hence  plays an important role in giving an observable magnitude of $T_c$ in \figdisp{Fig-1-Tc} and  \figdisp{Figure-Tc-Second}. 
 \label{Fig-Jeff}  }
 \end{figure}

In \figdisp{Fig-Jeff} we illustrate the role of the feedback enhancement of the exchange $J$ due to the  background spectral function discussed in \disp{Jeff}.
 \begin{figure}[htbp]
\centering
\includegraphics[width=.63\columnwidth]{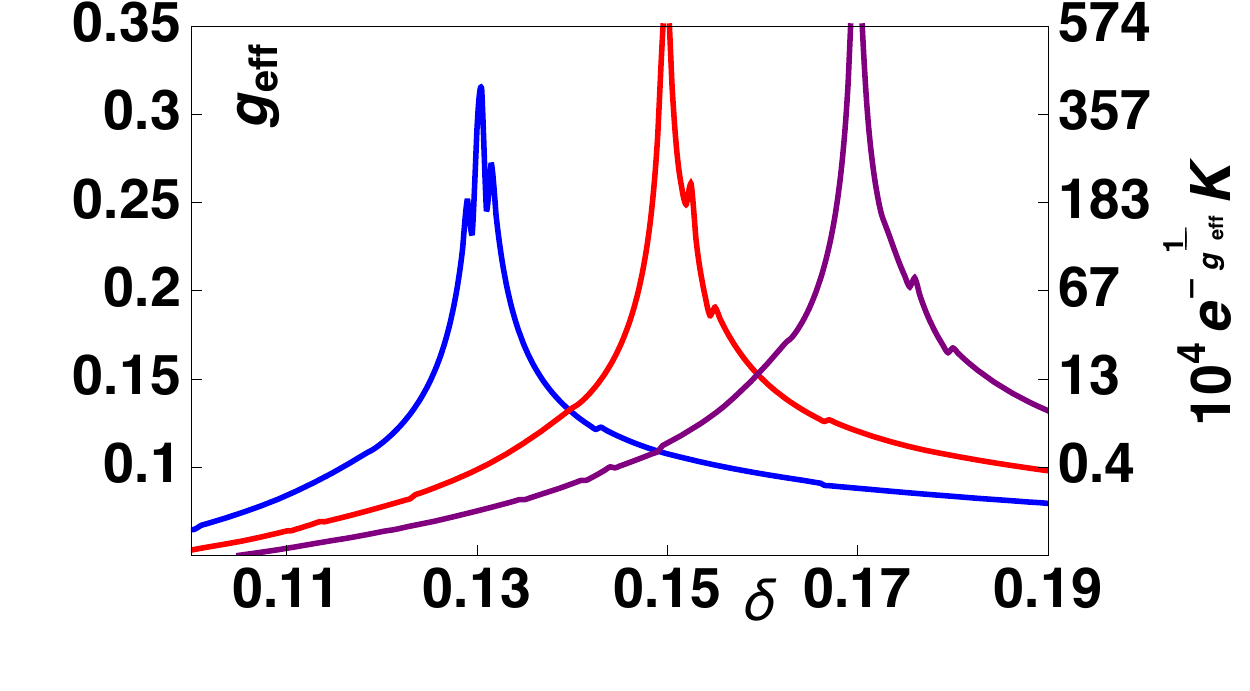}
 \caption{\footnotesize
The effective superconducting coupling $g_{eff}$ (\disp{geff})  for the three  curves in \figdisp{Figure-Tc-Second}, with  parameter sets- (i) (red) $t'/t$$=-0.159$, $t''/t$$=0.01$ with $\delta_{peak}=0.15$,  (ii) (blue) $t'/t$$=-0.137$, $t''/t=.01$  with $\delta_{peak}=0.13$ and (iii)(purple)
$t'/t$$=-0.181$, $t''/t$$=0.01$ with $\delta_{peak}=0.17$.  The drastic decrease of $T_c$ on both sides of the peak values in \figdisp{Figure-Tc-Second}  can be understood by referring to the the second y-axis, giving the approximate temperature scale $T_c^{appx}= 10^4 \times e^{- \frac{1}{g_{eff}}}$ K.  \label{Fig-geff}  }
 \end{figure}
For each set of parameters, there is a  density  region where both the  DOS at the fermi energy and the averaged d-wavefunction are enhanced, and the confluence directly enhances $J_{eff}$.  In turn this is reflected in the superconducting coupling $g_{eff}$. In \figdisp{Fig-geff} we see how  the confluence of   enhancements in the DOS and in the d-wavefunction $\ppsi$, further boosts the superconducting coupling  $g_{eff}$ and offsets to some extent the suppression due to a small magnitude of $Z$, as seen in \disp{geff}. As a result of this competition $T_c$ turns out to be in the observable range. The additional y-axis in \figdisp{Fig-geff} translates the superconducting  coupling $g_{eff}$ to  an order of magnitude type transition temperature $T^{appx}_c=  e^{-1/g_{eff}} \times 10^4 \; K$. This scale  helps us to understand why $T_c$ falls off so rapidly when $\delta$ increases beyond the peak value where the coupling $g_{eff}$ falls  below $\sim 0.12$, thereby rapidly suppressing $T_c$. 

\section{Conclusions \label{Conclusions}}

This work presents  a new methodology for treating extremely correlated superconductors. 
The exact equations of the normal and anomalous Greens  functions in the superconductor  are derived. These are further expanded in powers of a control parameter $\lambda$ related to the density of double occupancy, and the second order equations are given in Eqs.~(\ref{new-1},\ref{new-2},\ref{new-3}), together with the self consistency conditions  Eqs.~(\ref{cstarf},\ref{gap-1}).  A further simplification  is possible for $T$$\sim$$T_c$ where  the anomalous terms are small. This  leads to a tractable condition for $T_c$ given in \disp{Tc20}, expressed   in terms of the electron spectral function. Further  analysis uses  a model spectral function \disp{FLT}, which is simple enough to yield  an explicit expression for $T_c$ in \disp{Tc}. More elaborate calculations should be feasible upon the availability of reliable   spectral functions, when one may  directly  solve \disp{Tc1}.

\sr{ Our calculation delineates  the regime of parameters where superconductivity is possible in the \tJ model  within the  ECFL theory. This regime turns out to be  quite constrained.  The calculation highlights the requirement of a substantial magnitude of the  d-wavefunction average and  the DOS at the fermi energy. It
 shows that $T_c$ is maximal at a density where $n(\mu_0)$, the bare DOS is peaked, and is co-located with the peak of the fermi surface average of the d-wavefunction \disp{D-wave} (inset  \figdisp{Figure-Tc-Second}). The latter aspect  is understandable, since  the passing of the Fermi level energy dispersion   through the zone boundary points $\{\pm \pi,0\}, \{0,\pm \pi\} $, promotes a peak in the  DOS, and also leads to the maximization of $(\cos p_x-\cos p_y)^2$.
 The prediction  of a correlation between the peak in $T_c$ with a peak in the d-wavefunction average is testable, since the latter is amenable to measurement using angle resolved photoemission.}

  In the approximation used here, the maximum $T_c$ is nominally unbounded  in a narrow density range here  due   to the logarithmic singularity of the DOS. It  is expected to be  cutoff to a finite value of ${\cal O }(10^2\mbox{K})$ due to   a more exact  integration over energies, when using a reliable spectral function, in the place of the model used here. Such an integration would also  supersede the Gor'kov-type  approximation of expanding around the fermi surface ($\int d\epsilon \; n(\epsilon) \sim n(\mu_0) \int d\epsilon$) employed here, thereby flattening out the sharp peak into a smoother shape.  Finally this mean-field description of the superconductor is expected to be corrected by fluctuations  of the phase, in a strictly two dimensional case, and  by interlayer coupling, in the physically realistic case of a three dimensional system of weakly coupled layers.
 
In conclusion this work contains the essential outline of a new formalism to treat superconducting states of models with extremely strong correlations, such as the \tJ model. 
 A transparent  calculation within a low order approximation is presented here. It   demonstrates that the exchange energy $J$  can indeed provide  the fundamental binding force between electrons forming   Cooper pairs. It leads to superconductivity with $T_c$'s of ${\cal O }(10^2\mbox{K})$, in  a finite  range of densities located away from the insulator, as also experimentally  found    in   cuprate superconductors.

 \section{Acknowledgements:} 
  
 The work at UCSC was supported by the US Department of Energy (DOE), Office of Science, Basic Energy Sciences (BES), under Award No. DE-FG02-06ER46319.




 \end{document}